\documentclass{article}

\usepackage{amssymb}
\usepackage{amsmath}
\usepackage{bm}
\usepackage{subcaption}
\usepackage{caption}
\usepackage{multirow}
\usepackage{array}
\usepackage{arxiv}

\usepackage[utf8]{inputenc} 
\usepackage[T1]{fontenc}    
\usepackage{hyperref}       
\usepackage{url}            
\usepackage{booktabs}       
\usepackage{amsfonts}       
\usepackage{nicefrac}       
\usepackage{microtype}      
\usepackage{lipsum}
\usepackage{graphicx}
\graphicspath{ {./images/} }

\title{Signal-based online acceleration and strain data fusion using B-splines and Kalman filter for full-field dynamic displacement estimation}

\author{
 Aniruddha Das \\
  Department of Civil and Environmental Engineering\\
  Rice University\\
  Houston, Texas, U.S.A. - 77005\\
  \texttt{aniruddha.das@rice.edu} \\
  \And
 Ashish Pal \\
  Department of Civil and Environmental Engineering\\
  Rice University\\
  Houston, Texas, U.S.A. - 77005\\
  \texttt{ashish.pal@rice.edu} \\
  \And
 Satish Nagarajaiah \\
  Department of Civil and Environmental Engineering\\
  Department of Mechanical Engineering\\
  Rice University\\
  Houston, Texas, U.S.A. - 77005\\
  \texttt{satish.nagarajaiah@rice.edu} \\
  \And
 Mohamed Sajeer M \\
  Department of Civil Engineering\\
  Indian Institute of Technology Kanpur\\
  Kanpur, Uttar Pradesh, India - 208016\\
  \texttt{sajeermodavan@gmail.com} \\
  \And
 Suparno Mukhopadhyay \\
  Department of Civil Engineering\\
  Indian Institute of Technology Kanpur\\
  Kanpur, Uttar Pradesh, India - 208016\\
  \texttt{suparno@iitk.ac.in}
}

\begin{document}
\maketitle
\begin{abstract}
Displacement plays a crucial role in structural health monitoring (SHM) and damage detection of structural systems subjected to dynamic loads. However, due to the inconvenience associated with the direct measurement of displacement during dynamic loading and the high cost of displacement sensors, the use of displacement measurements often gets restricted. In recent years, indirect estimation of displacement from acceleration and strain data has gained popularity. Several researchers have developed data fusion techniques to estimate displacement from acceleration and strain data. However, existing data fusion techniques mostly rely on system properties like mode shapes or finite element models and require accurate knowledge about the system for successful implementation. Hence, they have the inherent limitation of their applicability being restricted to relatively simple structures where such information is easily available. In this article, B-spline basis functions have been used to formulate a Kalman filter-based algorithm for acceleration and strain data fusion using only elementary information about the system, such as the geometry and boundary conditions, which is the major advantage of this method. Also, the proposed algorithm enables us to monitor the full-field displacement of the system online with only a limited number of sensors. The method has been validated on a numerically generated dataset from the finite element model of a tapered beam subjected to dynamic excitation. Later, the proposed data fusion technique was applied to an experimental benchmark test of a wind turbine blade under dynamic load to estimate the displacement time history. In both cases, the reconstructed displacement from strain and acceleration was found to match well with the response from the FE model.
\end{abstract}


\section{Introduction}
\label{sec:introduction}
Structural infrastructure systems often experience extreme dynamic loads from natural hazards like earthquakes, hurricanes, and wind during service life. Regular maintenance and prompt identification of damage in structures reduce repair costs, while decision-making becomes easier. Therefore, continuous condition assessment and damage detection for structural health monitoring of these structures under operational conditions become crucial to ensure structural integrity and avoid major structural failures and subsequent disruption to life and the economy. Regular monitoring of dynamic parameters of the structure becomes important for condition assessment and SHM. Popularly, acceleration and strain data have been used to perform SHM of structures. However, the use of displacement data is most suited for identifying and localizing damage as it contains both global and location information and relates well with other structural properties of the system. Several researchers have used dynamic displacement for damage detection in civil engineering structures \cite{nagarajaiah2016structural, nagarajaiah2017modeling, yang2016harnessing, zhu2019damage, wang2021research}. A neural network-based direct parametric identification and damage detection methodology using dynamic displacement measurement data was proposed by Xu et al. \cite{xu2012damage}. A damage detection algorithm for bridges under moving loads was developed by Sun et al. \cite{sun2016damage} by decomposing the deflection into dynamic and quasi-static components.

Displacement estimation techniques in structural systems under dynamic loads can be broadly divided into direct and indirect methods. Direct displacement measurement techniques include the use of linear variable differential transducer or LVDT \cite{moreu2015dynamic, santhosh2017online} and Laser Doppler Vibrometer or LDV \cite{rossi2002comparison, garg2019noncontact, malekjafarian2018feasibility}, global position systems (GPSs) \cite{im2013summary, yi2013recent, de2019vibration, jo2013feasibility}, and computer vision-based camera systems \cite{fukuda2013vision, santos2016vision, feng2018computer, gao2023structural}. The LVDT is a contact-based displacement measurement device that requires firm support near the target point, which is not feasible for monitoring large structures with complex dynamic loading. The LDV is a reference-based, non-contact type of sensor that works on the the principle of incident and reflected laser beams to and from the target point. It requires a clear path between the sensor and the target point, which reflects light perpendicularly. Due to installation issues and the high cost of the set-up, LVDTs and LDVs are hardly used for displacement measurements in dynamic systems. GPS and computer-vision-based displacement measurement techniques gave cheaper and reference-free alternatives. However, the GPS-based methods lack the necessary measurement requirements for SHM and damage detection due to low sampling rates and less accuracy. Occlusion due to high-rise buildings and environmental conditions affect data collection; hence, GPS-based measurements are not reliable. Recently, camera-based systems have been developed to measure displacements in complex structures, and they give quite accurate measurements \cite{dong2020structural, bhowmick2020measurement, luan2021extracting, jana2022physics, jana2023data}. However, cameras are difficult to install in tall structures, and they require good visual conditions to operate accurately. Also, they have a low sampling rate due to hardware limitations, which pose practical difficulties in dynamic displacement tracking.

Indirect estimation of displacements in dynamic systems involves displacement reconstruction from other measured quantities like acceleration \cite{lee2010design, sabatini2015fourier}, and strain \cite{metje2008optical, chan2009vertical, rapp2009displacement}. The high-frequency content gets captured accurately in acceleration signals, and therefore, numerical double integration of measured acceleration to estimate displacement has gained popularity \cite{hudson1979reading}. However, error accumulation from double integration often causes a baseline drift, particularly in the case of online estimation of displacement in long-term SHM \cite{thong2004numerical}. Due to limitations of displacement estimation from acceleration data, several researchers have made efforts to estimate dynamic displacement from other easily measurable quantities like strain using the strain-displacement relations, such as strain-based mode shapes and mode superposition \cite{wang2014strain, shin2012estimation, davis1996shape}. However, these techniques are subject to modal analysis errors and require detailed system information for successful implementation.

Recently, many researchers have combined the direct and indirect methods of displacement measurement to develop data fusion techniques for displacement estimation. Kalman filters have been used to fuse high-frequency acceleration and low-frequency displacement data (from GPS or camera-based methods) \cite{smyth2007multi, kim2014autonomous, kim2017dynamic, pal2024data}. The application of such techniques gets restricted due to visibility and installation issues in displacement measurement via GPS and camera-based systems. Therefore, displacement estimation techniques by fusion of strain and acceleration data have gained popularity in recent years. Displacement of a simply-supported bridge was estimated by Park et al. \cite{park2013displacement, park2014wireless, park2018visual} by fusing acceleration and strain-derived displacement through a finite impulse response (FIR) filter \cite{lee2010design}. The strain-based displacement was used to regularize the displacement obtained by integrating acceleration, thereby preventing signal drift. However, the authors used sinusoidal mode shapes proposed by  Shin et al. \cite{shin2012estimation} for modal mapping, and hence, the technique is restricted to simple bridges. This work was further extended by Cho et al. \cite{cho2015displacement} to more complex types of bridges, like truss bridges using finite element (FE) based analytical mode shapes. Later, a Kalman filter-based method for combining acceleration and mode shape-based displacement was proposed by Cho et al. \cite{cho2016reference} to obtain dynamic displacement of complex bridges. This technique also uses analytical (sinusoidal) mode shapes and is not suitable for complex structures where developing an FE model is difficult without detailed knowledge about the system. A multi-rate Kalman filter-based displacement estimation technique using acceleration and strain data was proposed by Zhu et al. \cite{zhu2020multi} for super-tall structures. It is based on the geometrical deformation of the beam-like structures, and therefore, its application is limited to specific 1D beam-like structures. Most recently, another technique for the fusion of acceleration and strain data has been developed by  Ren et al. \cite{ren2022theoretical}, Ji et al. \cite{ji2023deformation}, which is based on the identification of the strain mode shapes using stochastic subspace identification (SSI). It proposes Kalman filter-based data fusion using acceleration as input in process equation and strain-derived displacement as a measurement for displacement estimation. However, the identification of strain-mode shapes from dynamic data is often difficult and may again be prone to errors, and hence, it restricts the applicability of this method for real-world structures.

In this study, a novel Kalman filter-based data fusion technique has been proposed for displacement estimation in dynamic systems using data fusion of measured acceleration and strain. In Section \ref{sec:Theory}, a detailed formulation for the proposed method has been presented by using B-spline basis functions. In the Kalman filter equations, the acceleration signal is used as input in the process equation, and flexural strain data is used as the measurement. The proposed method is solely based on signal processing and only uses very elementary system information like boundary conditions and the distance of the neutral axis from the surface at the strain gauge locations. Therefore, the proposed method has a versatile range of applicability, even for complex structures, where the development of FE model or mode shape information is difficult to obtain. In Section \ref{sec:NumericalValidation}, the proposed data fusion technique has been numerically validated on two cases of a tapered cantilever beam problem involving collocated and non-collocated sensors. Thereafter, in Section \ref{sec:ExpValidation}, the experimental validation of our method has been demonstrated on an experimental benchmark for small-scale dynamic testing of a wind turbine blade.

\section{Theoretical Formulation}
\label{sec:Theory}
\begin{figure}
    \centering
    \includegraphics[width=0.9\linewidth]{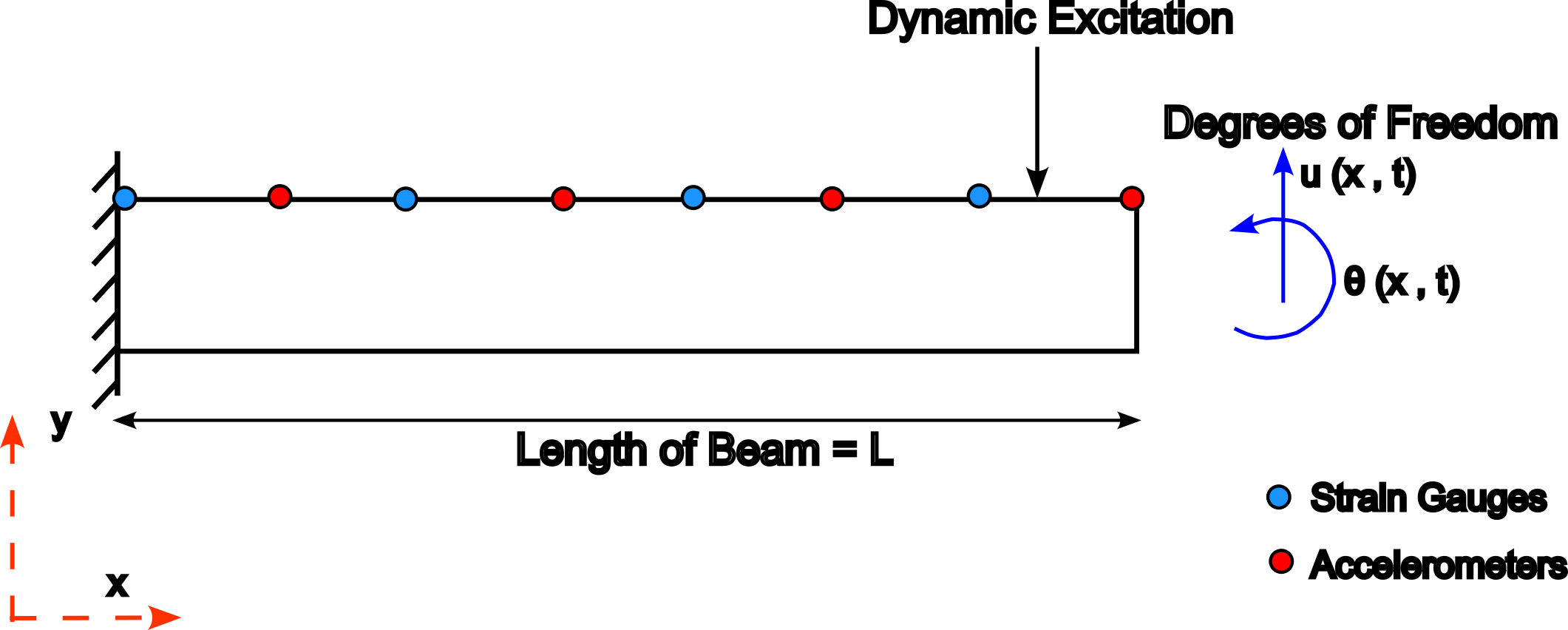}
    \caption{Cantilever beam under dynamic excitation}
    \label{fig:BeamDescription}
\end{figure}

Let us consider a tapered cantilever beam of length $L$ and depth $h(x)$, as shown in Figure \ref{fig:BeamDescription}. The beam is excited by an unknown dynamic flexural load at any location along the length. The FE model is developed by dividing the length of the beam into $n_{e} (= n-1)$ number of linear elements of equal lengths with two degrees of freedom, viz., the vertical displacement $u(x,t)$ and slope $u^{'}(x,t) = \theta (x,t)$ at each of the $n$ nodes.

Let the acceleration and strain time histories be recorded at random points along the beam. It may be noted here that the recorded system response data are discrete in both time and space. As we know, acceleration is the second derivative of displacement with respect to time, while flexural strain is associated with the second spatial derivative of displacement. In order to fuse the discrete acceleration and flexural strain data, it is crucial to express the system response as a continuous analytical function over time and space domain so that both the physical quantities (acceleration and flexural strain) can be expressed analytically. Therefore, the B-spline basis functions have been used in this study to express the dynamic displacement of the system analytically and, thus, also avoid the use of any major system information, such as analytical mode shapes or the FE model of the system.

\subsection{B-spline basis functions}
\label{sec:B-spline}
\begin{figure}
    \centering
    \includegraphics[width=1.0\linewidth]{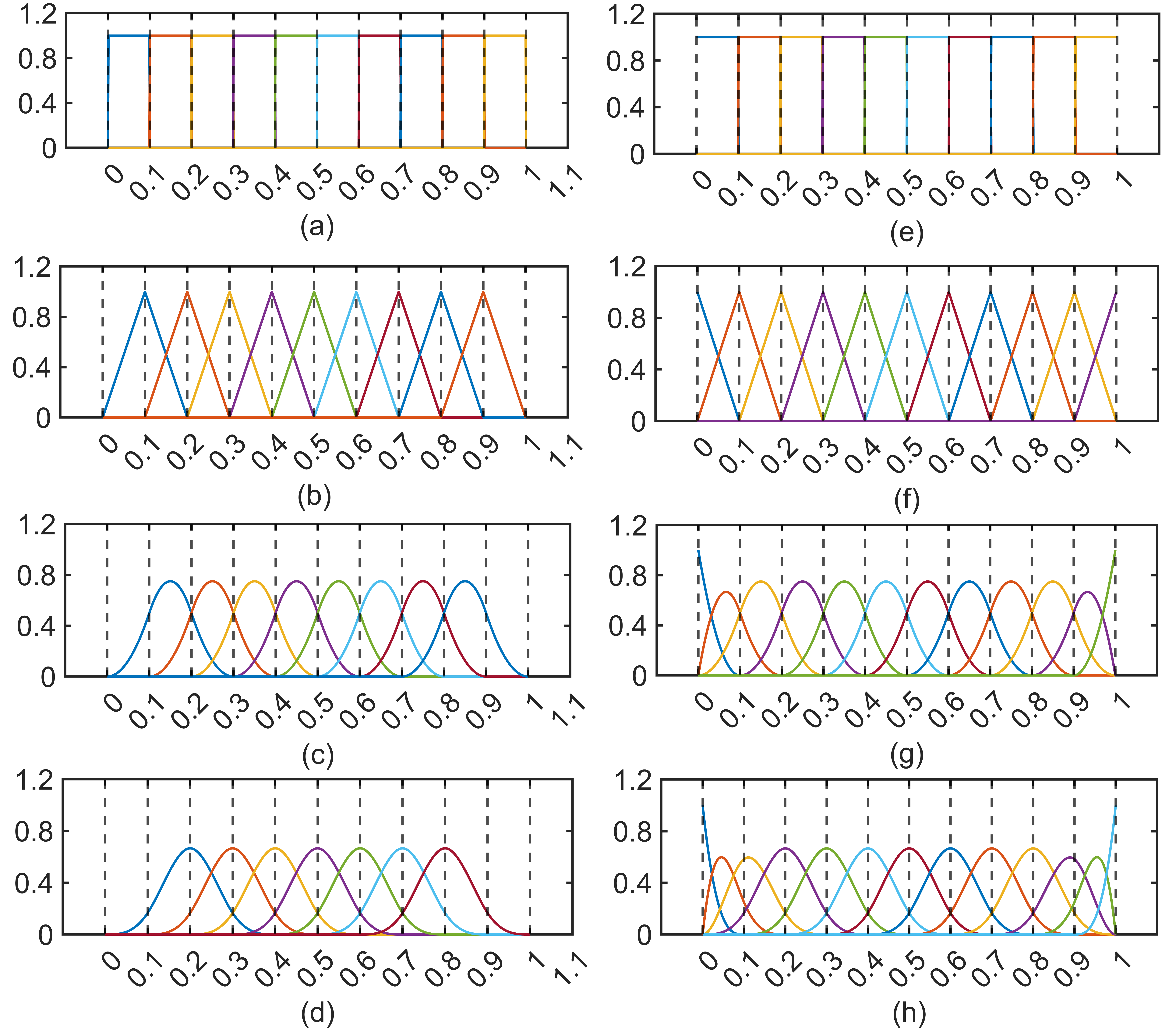}
    \caption{B-spline basis functions: (a)-(d) closed-ended splines, (e)-(f) open-ended splines}
    \label{fig:SplineExample}
\end{figure}

B-spline basis functions are piecewise polynomial functions of any specific order over a domain divided into a grid of non-decreasing knots. Each individual basis function is non-zero only over a particular interval in the domain, while zero otherwise, thereby demonstrating its property of compact (local) support. In one dimension, these basis functions are univariate polynomials that can be defined over a domain divided into equal or unequal intervals with $k$ number of knots, including the boundaries. The Cox-de-Boor \cite{de1978practical} algorithm can be used recursively to obtain the 1D B-spline basis functions up to the desired order from the lower-order basis functions. A sequence of B-splines up to order four, defined over the domain $[0, 1]$, is shown in Figure \ref{fig:SplineExample}, with 11 equidistant knots marked by the dashed vertical lines. B-spline functions may be defined as close-ended (Figure \ref{fig:SplineExample} (a)-(d)) or open-ended (Figure \ref{fig:SplineExample} (e) - (h)) at the supports, depending on the requirement. In the current study, a set of closed-ended B-spline basis functions of order four (degree three) has been adopted to write the displacement function as a continuous function of time and space.

\subsection{Formulation for continuous time and space}
\label{sec:ContStateSpaceFormulation}
Let the dynamic displacement field of the beam under dynamic excitation be given by $u(x,t)$. The function $u(x,t)$ can be written in terms of $m$ basis functions as

\begin{equation}
    \label{eq:cont-disp}
    {u(x,t)=\sum_{i=1}^{m} {\phi_{i}(x)\lambda_{i}(t)}} = \boldsymbol{\phi}(x) \boldsymbol{\lambda}(t),
\end{equation}
where $\lambda_{i}(t)$ denotes the contribution of the $i^{th}$ B-spline basis function, $\phi_{i}(x)$, to the displacement at any point $x$ along the length of the beam at any time instant $t(\geq0)$ and $m$ is the number of basis functions used in the formulation. In the vectorial representation of Equation \eqref{eq:cont-disp}, $\bm{\phi}(x) = \begin{Bmatrix} \phi_{1}(x) & \phi_{2}(x) & ... & \phi_{m}(x) \end{Bmatrix}$ and $\boldsymbol{\lambda}(t) = \begin{Bmatrix} \lambda_{1}(t) & \lambda_{2}(t) & ... & \lambda_{m}(t) \end{Bmatrix}^{T}$. The acceleration at any point $x$ along the beam can be derived as

\begin{equation}
    \label{eq:cont-acc}
    {\ddot{u}(x,t)=\frac{\partial^2 u}{\partial t^2}=\sum_{i=1}^{m} {\phi_{i}(x)\ddot{\lambda}_{i}(t)}} = \bm{\phi}(x) \bm{\ddot{\lambda}}(t).
\end{equation}

 Further, the expression for flexural strain $\varepsilon(x,t)$ at any point $x$, at a distance of $z(x)$ from the neutral axis, can be written as:
 
 \begin{equation}
 \label{eq:cont-strain}
    {\varepsilon(x,t)=-z(x)\frac{\partial^2 u}{\partial x^2}= -z(x)\sum_{i=1}^{m}{\phi^{\prime\prime}_{i}(x)\lambda_{i}(t)}}
     = -z(x) \boldsymbol{\phi^{''}}(x) \boldsymbol{\lambda}(t).
\end{equation}
It may be noted here that the dynamic displacement $u(x,t)$ can be obtained from the flexural strain by integrating it twice with respect to space as:
\begin{equation}
\label{eq:cont-DispFromStrain}
    u(x,t) = \sum_{i=1}^{m}{\lambda}_{i}(t) \iint \phi^{\prime\prime}_{i}(x){dx}\cdot{dx} + B(t)x + C(t),
\end{equation}
where B(t) and C(t) are independent of space and can be evaluated from the boundary conditions. In the case of the cantilever beam problem, since the beam is fixed at $x=0$, both $B(t)$ and $C(t)$ will be equal to $0 \:{\forall \, t\geq 0}$.

The time-varying parameters $\lambda_{i}(t)$ in the above formulation can be determined by setting up Equations \eqref{eq:cont-disp}, \eqref{eq:cont-acc}, \eqref{eq:cont-strain} and \eqref{eq:cont-DispFromStrain} in a state-space formulation and solving it using a Kalman Filter-based iterative method from acceleration and strain data. Therefore, the process equation or the time update equation for the Kalman filter can be written using Equations \eqref{eq:cont-disp} and \eqref{eq:cont-acc} as
\begin{equation}
    \begin{bmatrix}
        \boldsymbol{\dot{\lambda}}(t) \\
        \boldsymbol{\ddot{\lambda}}(t)
    \end{bmatrix} = 
    \underbrace {\begin{bmatrix}
        \mathbf{0} & \mathbf{I} \\
        \mathbf{0} & \mathbf{0}
    \end{bmatrix}}_{\mathbf{A_{c}}}
    \begin{bmatrix}
        \boldsymbol{\lambda}(t) \\
        \boldsymbol{\dot{\lambda}}(t)
    \end{bmatrix} + 
    \underbrace {\begin{bmatrix}
        \mathbf{\mathbf{0}} \\
        \mathbf{\mathbf{I}}
    \end{bmatrix} \boldsymbol{\phi}(x)^{\dagger}}_{\mathbf{B_{c}}} \bm{\ddot{u}}(x,t) + 
    \begin{bmatrix}
        \mathbf{\mathbf{0}} \\
        \mathbf{\mathbf{I}}
    \end{bmatrix} \boldsymbol{\phi}(x)^{\dagger} \boldsymbol{\eta_{acc}},
\end{equation}
and, similarly, from Equations \eqref{eq:cont-disp} \eqref{eq:cont-strain} and \eqref{eq:cont-DispFromStrain}, the measurement update equation can be written using as
\begin{equation}
\label{eq:Cont-Measurement}
    \begin{Bmatrix} \boldsymbol{\varepsilon}(x,t) \\ \boldsymbol{u}(\boldsymbol{\bar x_{u}},t) \\ \boldsymbol{u}^{'}(\boldsymbol{\bar x_{u^{'}}},t) \end{Bmatrix} = 
    \underbrace{ \begin{bmatrix}
        -z(x) {\boldsymbol{\phi^{''}} (x)} & \mathbf{0} \\
        {\boldsymbol{\phi}(\boldsymbol{\bar x_{u}})} & \mathbf{0} \\
        {\boldsymbol{\phi^{'}}(\boldsymbol{\bar x_{u^{'}}})} & \mathbf{0}
    \end{bmatrix}}_{\mathbf{C_{c}}}
    \begin{bmatrix}
        \boldsymbol{\lambda}(t) \\
        \boldsymbol{\dot{\lambda}}(t)
    \end{bmatrix} + 
    \begin{bmatrix} \mathbf{I} \\ \mathbf{0} \\ \mathbf{0} \end{bmatrix} \boldsymbol{\eta_{strain}}
\end{equation}
where $\begin{bmatrix} \boldsymbol{\lambda}(t) \\ \boldsymbol{\dot{\lambda}}(t) \end{bmatrix}$ denote the state space vector, and $(\dagger)$ operator represents the Moore–Penrose inverse or pseudo-inverse of a matrix. Here, $\mathbf{A_c}$, $\mathbf{B_c}$, and $\mathbf{C_c}$ are the state transition matrix, the control input matrix, and the state observation matrix respectively; while $\boldsymbol{\eta_{acc}}$ and $\boldsymbol{\eta_{strain}}$ denote the noises associated with the acceleration and strain measurements respectively. The noises in each acceleration and strain measurement are assumed to be zero mean Gaussian processes with standard deviations as $\mathbf{q_{acc}}$ and $\mathbf{r_{strain}}$, respectively. Here, in Equation \eqref{eq:Cont-Measurement}, we incorporate the boundary conditions of the problem in accordance with Equation \eqref{eq:cont-DispFromStrain}. The vectors $\boldsymbol{u}(\boldsymbol{\bar x_{u}},t)$ and $\boldsymbol{u}^{'}(\boldsymbol{\bar x_{u^{'}}},t)$ denote the displacement and slope boundary conditions of the problem given at locations $\boldsymbol{\bar x_{u}}$ and $\boldsymbol{\bar x_{u^{'}}}$ respectively.

The above formulation can be written in compact form as follows:
\begin{equation}
    \boldsymbol{\dot{\Lambda}}(t) = \mathbf{A_{c}} \boldsymbol{\Lambda}(t) + \mathbf{B_{c}}\bm{\ddot{u}}(x,t) + \mathbf{w}(t)
\end{equation}
and
\begin{equation}
    \boldsymbol{\Gamma}(t) = \mathbf{C_{c}} \boldsymbol{\Lambda}(t) + \mathbf{v}(t)
\end{equation}
where the state vector is denoted by $\boldsymbol{\Lambda}(t)$ while the $\mathbf{w}(t)$ and $\mathbf{v}(t)$ are zeros mean Gaussian processes with their respective co-variances $\mathbf{Q}$ and $\mathbf{R}$, as described below:
\begin{equation}
    \mathbf{Q} = \begin{bmatrix}
        \mathbf{0} & \mathbf{0} \\
        \mathbf{0} & \begin{Bmatrix} \boldsymbol{\phi}(x)^{\dagger} diag(\mathbf{q_{acc}}) \end{Bmatrix}
        \begin{Bmatrix} \boldsymbol{\phi}(x)^{\dagger} diag(\mathbf{q_{acc}}) \end{Bmatrix} ^{T}
    \end{bmatrix} = 
    \begin{bmatrix}
        \mathbf{0} & \mathbf{0} \\
        \mathbf{0} & \widehat{\mathbf{Q}}
    \end{bmatrix},
\end{equation}
and
\begin{equation}
    \mathbf{R} = diag\begin{bmatrix}
        \mathbf{r_{strain}} & \mathbf{0} & \mathbf{0}
    \end{bmatrix}^2 =
    \begin{bmatrix}
        r_{strain,1}^{2} & & & \mathbf{0} & \mathbf{0} \\
        & \ddots & & \vdots & \vdots \\
        & & r_{strain,q}^{2} & \mathbf{0} & \mathbf{0} \\
        \mathbf{0} & ... & \mathbf{0} & \mathbf{0} & \mathbf{0} \\
        \mathbf{0} & ... & \mathbf{0} & \mathbf{0} & \mathbf{0} \\
    \end{bmatrix}.
\end{equation}

The above set of equations explicitly describes the relation between the states, measurements, and associated noise vectors. Here, the acceleration data is used as input in the process or time update equation, and the strain data is observed measurement in the measurement equation. Moreover, the boundary condition has been carefully introduced into the formulation by inserting $u(0,t)$ and $u^{'}(0,t)$ in the measurement equation corresponding to the fixed support of the cantilever beam. In general, the boundary conditions can be easily incorporated into the measurement equation.

\subsection{Discrete-time state-space formulation}
\label{sec:DisFormulation}
For our tapered cantilever beam problem, the acceleration and strain measurements obtained from sensors are available only at discrete points along the length of the beam. Therefore, the above formulation in continuous time and space has to be adjusted for the discrete time and space domain. Let $p$ number of accelerometers and $q$ number of strain gauges be installed at different locations along the length of the beam, such that both $\begin{Bmatrix} p,q \geq m\end{Bmatrix}$ (as shown in Figure \ref{fig:BeamDescription}). Here, we first adjust the Equations \eqref{eq:cont-disp}, \eqref{eq:cont-acc} and \eqref{eq:cont-strain} for discrete-space domain and then demonstrate how the discrete-time state space formulation can be derived.

Using Equation $\eqref{eq:cont-disp}$, the displacements at any $n$ number of locations (denoted by $\boldsymbol{x}_{n \times 1}$) along the length of the beam at any particular time instant $t$ can be written as
\begin{equation}
    \label{eq:dis-disp}
    \boldsymbol{u} (\boldsymbol{x}_{n \times 1}, t) = \begin{Bmatrix} u(x_{1},t) \\ : \\ u(x_{k},t) \\ : \\ u(x_{n},t) \end{Bmatrix} =
    \begin{Bmatrix} {\sum_{i=1}^{m}
    \phi_{i}(x_{k}) \lambda_{i}
    (t)} \mid k = {1,2,...,n}
    \end{Bmatrix}  = 
    \begin{bmatrix} \boldsymbol{\Phi} \end{bmatrix}_{n \times m} \boldsymbol{\lambda}_{m \times 1}(t).
\end{equation}
where $\begin{bmatrix} \boldsymbol{\Phi} \end{bmatrix}_{n \times m}$ denotes the matrix of $m$ number of B-spline basis functions defined at $n$ points in the domain. Let the matrices of B-spline functions defined at the accelerometer and strain gauge locations be denoted by $\begin{bmatrix} \boldsymbol{\Phi}_{acc} \end{bmatrix}_{p \times m}$ and $\begin{bmatrix} \boldsymbol{\Phi}_{strain} \end{bmatrix}_{q \times m}$, where $p$ and $q$ are the number of accelerometers and strain gauges respectively. Therefore, the expression for acceleration at the accelerometer locations can be written as:
\begin{equation}
\label{eq:dis-acc}
    \boldsymbol{\ddot u} (\boldsymbol{x}_{p \times 1}, t) = \begin{bmatrix} \boldsymbol{\Phi}_{acc} \end{bmatrix}_{p \times m} \boldsymbol{\ddot \lambda}_{m \times 1}(t),
\end{equation}
and the flexural strain at the location of the strain gauges can be written as:
\begin{equation}
\label{eq:dis-strain}
    \boldsymbol{\varepsilon}_{q \times 1} (t) =  - diag\begin {bmatrix} \boldsymbol{z}_{q \times 1} \end{bmatrix}_{q \times q} \boldsymbol{u^{''}} (\boldsymbol{x}_{q \times 1}, t) =
    - diag \begin{bmatrix} \boldsymbol{z}_{q \times 1} \end{bmatrix}_{q \times q} \begin{bmatrix} \boldsymbol{\Phi^{''}}_{strain} \end{bmatrix}_{q \times m} \boldsymbol{\lambda}_{m \times 1}(t),
\end{equation}
where $\begin{bmatrix} diag(\boldsymbol{z}_{q \times 1}) \end{bmatrix}_{q \times q}$ is a diagonal matrix with the depth of the neutral axis at each strain sensor location as the diagonal elements.

Now, the discrete-time state-space model can be achieved easily from the continuous-time state-space model derived in Section \ref{sec:ContStateSpaceFormulation}. The state transition matrix $A_{c}$ being a nilpotent matrix (i.e., $\mathbf{A_{c}}^{2} = \mathbf{0}$), the discrete state transition matrix can be written as:
\begin{equation}
    \mathbf{A_{d}} = e^{\mathbf{A_{c}} \Delta t} = \mathbf{I} + \mathbf{A_{c}} \Delta t =
    \begin{bmatrix}
        \mathbf{I} & \Delta t \mathbf{I} \\
        \mathbf{0} & \mathbf{I}
    \end{bmatrix}_{2m\times 2m}
\end{equation}
and the discrete control matrix $\mathbf{B_{d}}$ can be derived as:
\begin{equation}
    \mathbf{B_{d}} = \int_{0}^{\Delta t} {e^{\mathbf{A_{c}} \tau} \mathbf{B_{c}} d\tau} = \mathbf{B_{c}} \Delta t + \frac{\mathbf{A_{c}} \mathbf{B_{c}} {\Delta t}^{2}}{2} = 
    \begin{bmatrix}
    \frac{\Delta t^{2}}{2} \mathbf{I} \\
        \Delta t \mathbf{I}        
    \end{bmatrix}_{2m\times m}
    \begin{bmatrix}
        \boldsymbol{\Phi}^{\dagger}_{acc}
    \end{bmatrix}_{m\times p}
\end{equation}
where $\Delta t$ is the sampling interval for the acceleration and strain sensors.
The  noise covariance matrices in discrete time and space domains have been derived as follows:
\begin{equation}
    \mathbf{Q_{d}} = \int_{0}^{\Delta t} e^{\mathbf{A_{c}} \tau} \mathbf{Q} e^{\mathbf{A_{c}}^{T} \tau} d\tau =
    \begin{bmatrix}
        \frac{\Delta t^{3}}{3} \mathbf{\widehat{Q}} & \frac{\Delta t^{2}}{2}\mathbf{\widehat{Q}} \\
        \frac{\Delta t^{2}}{2} \mathbf{\widehat{Q}} & \Delta t \mathbf{\widehat{Q}}
    \end{bmatrix}
\end{equation}
and
\begin{equation}
    \mathbf{R_{d}} = \frac{\mathbf{R}}{\Delta t}.
\end{equation}
Therefore, using Equations (11)-(17) in the discrete-time and space domain, the process equation can be written as:
\begin{equation}
    \begin{split}
    \begin{bmatrix}
        \boldsymbol{\lambda}_{m\times1}(k+1) \\
        \boldsymbol{\dot{\lambda}}_{m\times1}(k+1)
    \end{bmatrix} = 
    \underbrace {\begin{bmatrix}
        \mathbf{I}_{m\times m} & \Delta t \mathbf{I}_{m\times m} \\
        \mathbf{0}_{m\times m} & \mathbf{I}_{m\times m}
    \end{bmatrix}}_{\mathbf{A_{d}}}
    \begin{bmatrix}
        \boldsymbol{\lambda}_{m\times1}(k) \\
        \boldsymbol{\dot{\lambda}}_{m\times1}(k)
    \end{bmatrix} + 
    \underbrace {\begin{bmatrix}
    \frac{\Delta t^{2}}{2} \mathbf{I} \\
        \Delta t \mathbf{I}        
    \end{bmatrix}_{2m\times m} \begin{bmatrix} \boldsymbol{\Phi}^{\dagger}_{acc} \end{bmatrix}_{m\times p}}_{\mathbf{B_{d}}} \bm{\ddot{u}^{meas}}_{p\times 1} + \\
    \begin{bmatrix}
    \frac{\Delta t^{2}}{2} \mathbf{I} \\
        \Delta t \mathbf{I}        
    \end{bmatrix}_{2m\times m}
    \begin{bmatrix} \boldsymbol{\Phi}^{\dagger}_{acc} \end{bmatrix}_{m\times p} \boldsymbol{\eta_{acc}},
    \end{split}
\end{equation}
and the measurement equation as:
\begin{equation}
    \begin{Bmatrix} \boldsymbol{\varepsilon}^{meas}_{q\times 1}(k) \\ \boldsymbol{u}(\boldsymbol{x}_{\alpha},k) \\ \boldsymbol{u^{'}} (\boldsymbol{x}_{\beta},k) \end{Bmatrix} = 
    \underbrace{\begin{bmatrix}
        \begin{bmatrix} - diag \begin{bmatrix} \boldsymbol{z}_{q \times 1} \end{bmatrix}_{q \times q} \boldsymbol{\Phi}^{''} _{strain} \end{bmatrix}_{q\times m} & \mathbf{0}_{q\times m} \\
        {\boldsymbol{\Phi}_{\alpha \times m} (\boldsymbol{x}_{\alpha})} & \mathbf{0}_{\alpha \times m} \\
        {\boldsymbol{\Phi^{'}}_{\beta \times m} (\boldsymbol{x}_{\beta})} & \mathbf{0}_{\beta \times m}
    \end{bmatrix}}_{\mathbf{C_{d}}}
    \begin{bmatrix}
        \boldsymbol{\lambda}_{m \times 1}(k) \\
        \boldsymbol{\dot{\lambda}}_{m \times 1}(k)
    \end{bmatrix} + 
    \begin{bmatrix} \mathbf{I}_{q\times q} \\ \mathbf{0}_{\alpha \times 1} \\ \mathbf{0}_{\beta \times 1} \end{bmatrix} \boldsymbol{\eta _{strain}}.
\end{equation}
where the $\boldsymbol{u} (\boldsymbol{x}_{\alpha},k)$ and $\boldsymbol{u^{'}}(\boldsymbol{x}_{\beta},k)$ denote the vertical displacement and slope at $\alpha$ and $\beta$ number of locations on the beam at $k^{th}$ time instant and they can be determined based on the given boundary conditions of the problem.

In compact form, the process and measurement equations for discrete-time space formulation can be written as:
\begin{equation}
    \boldsymbol{\Lambda}(k+1) = \mathbf{A_{d}}\boldsymbol{\Lambda}(k) + \mathbf{B_{d}}\bm{\ddot{u}^{meas}}(k) + \mathbf{w_{d}}(k)
\end{equation}
and
\begin{equation}
    \boldsymbol{\Gamma}(k) = \mathbf{C_{d}}\boldsymbol{\Lambda}(k) + \mathbf{v_{d}}(k).
\end{equation}
The process and measurement equations formulated above describe the discrete-time Kalman filter algorithm that may be applied to accurately estimate the $\boldsymbol{\lambda}(k)$ at each time step. The time update, Kalman gain calculation, and measurement update steps in each iteration of the online Kalman filtering process have been described below:

Time update:
\begin{equation}
    \boldsymbol{\Lambda}(k+1 \mid k) = \mathbf{A_{d}}\boldsymbol{\Lambda}(k) + \mathbf{B_{d}} \bm{\ddot{u}^{meas}}(k)
\end{equation}
\begin{equation}
    \boldsymbol{\Theta}(k+1 \mid k) = \mathbf{A_{d}}\boldsymbol{\Theta}(k)\mathbf{A_{d}^{T}} + \mathbf{Q_{d}}
\end{equation}

Kalman gain:
\begin{equation}
    \mathbf{K_{g}}(k+1) = \boldsymbol{\Theta}(k+1 \mid k) \mathbf{C_{d}^{T}} \begin{bmatrix} \mathbf{C_{d}} \boldsymbol{\Theta}(k+1 \mid k)  \mathbf{C_{d}^{T}} + \mathbf{R_{d}} \end{bmatrix}^{-1}
\end{equation}

Measurement update:
\begin{equation}
    \boldsymbol{\Lambda}(k+1 \mid k+1) = \boldsymbol{\Lambda}(k+1 \mid k) + \mathbf{K_{g}}(k+1) \begin{bmatrix} \boldsymbol{\Gamma}(k+1) - \mathbf{C_{d}} \boldsymbol{\Lambda}(k+1 \mid k) \end{bmatrix}
\end{equation}
\begin{equation}
    \boldsymbol{\Theta}(k+1 \mid k+1) = \begin{bmatrix} \mathbf{I} - \mathbf{K_{g}}(k+1) \mathbf{C_{d}} \end{bmatrix} \boldsymbol{\Theta}(k+1 \mid k)
\end{equation}
Here, $\boldsymbol{\Theta}(.)$ denotes the covariance matrix of error associated with the time and measurement updates at each iteration.

\section{Numerical Validation}
\label{sec:NumericalValidation}
In this section, the proposed data fusion technique has been numerically validated for the tapered cantilever beam problem by considering two different sensor layouts. The beam is of length $L = 1.65 \text{ m}$ and width $b = 20 \text{ mm}$ and its depth varies from $h_{1} = 10 \text{ mm}$ to $h_{2} = 1 \text{ mm}$. The material considered for the beam is steel with a density of $7850 \text{ kg/m}^{3}$ and Young's modulus of $E = 2.1 \times 10^{11} \text{ N/m}^{2}$. It is fixed at one end, i.e., the displacement and slope at $x = 0$ are $u(0,t) = 0$ and $\dot{u}(0,t) = 0$ respectively and is excited by a dynamic load at the free end for a time period of $40 \text{ sec}$. Proportional damping or Rayleigh damping has been considered for this problem, assuming the damping ratios associated with the first and second modes are $\zeta_{1} = 3 \%$ and $\zeta_{2} = 4 \%$ respectively. In the following sections, the data fusion technique is first demonstrated by placing a pair of acceleration and strain sensors at the same locations (i.e., the acceleration and strain signals are measured at the same points along the length of the beam). Later, the same tapered cantilever beam but with non-collocated acceleration and strain sensors has been considered to demonstrate the efficiency of the proposed data fusion technique.

\subsection{Data-fusion for a cantilever beam with collocated sensors}
\label{sec:NumVaildCollocatedSensors}
\begin{figure}
    \centering
    \includegraphics[width=1.0\linewidth]{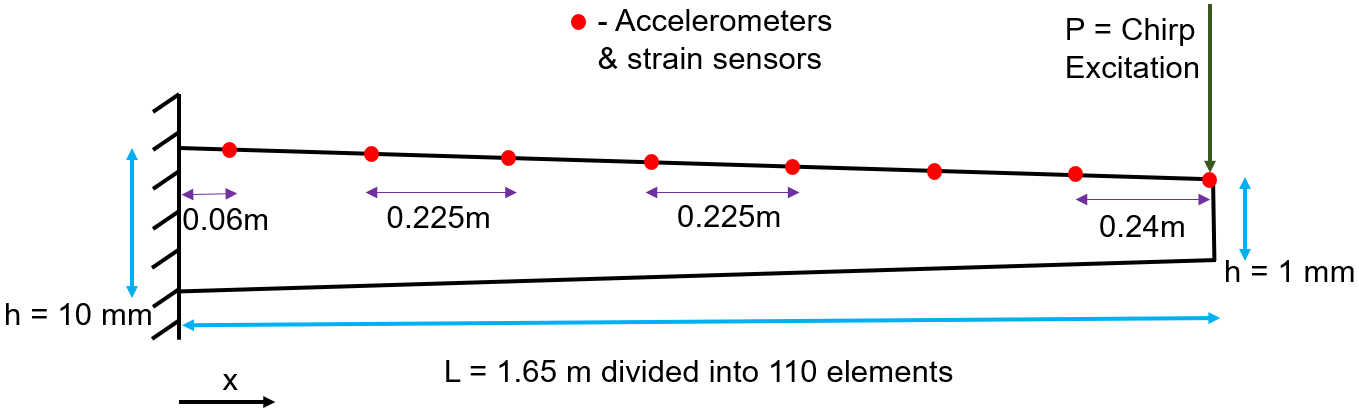}
    \caption{Tapered cantilever beam with collocated sensors under chirp excitation}
    \label{fig:CollocatedTaperedCantilever}
\end{figure}

The tapered cantilever beam problem with collocated acceleration and strain sensor pairs is shown in Figure \ref{fig:CollocatedTaperedCantilever}. It has eight pairs of accelerometers and strain gauges at the same eight locations, and the beam is subjected to chirp excitation of frequency ranging from $3-15 \text{ Hz}$, varying linearly, at the free end, as described in Figure \ref{fig:CollocatedTaperedCantilever}.

\begin{figure}
    \centering
    \includegraphics[width=1.0\linewidth]{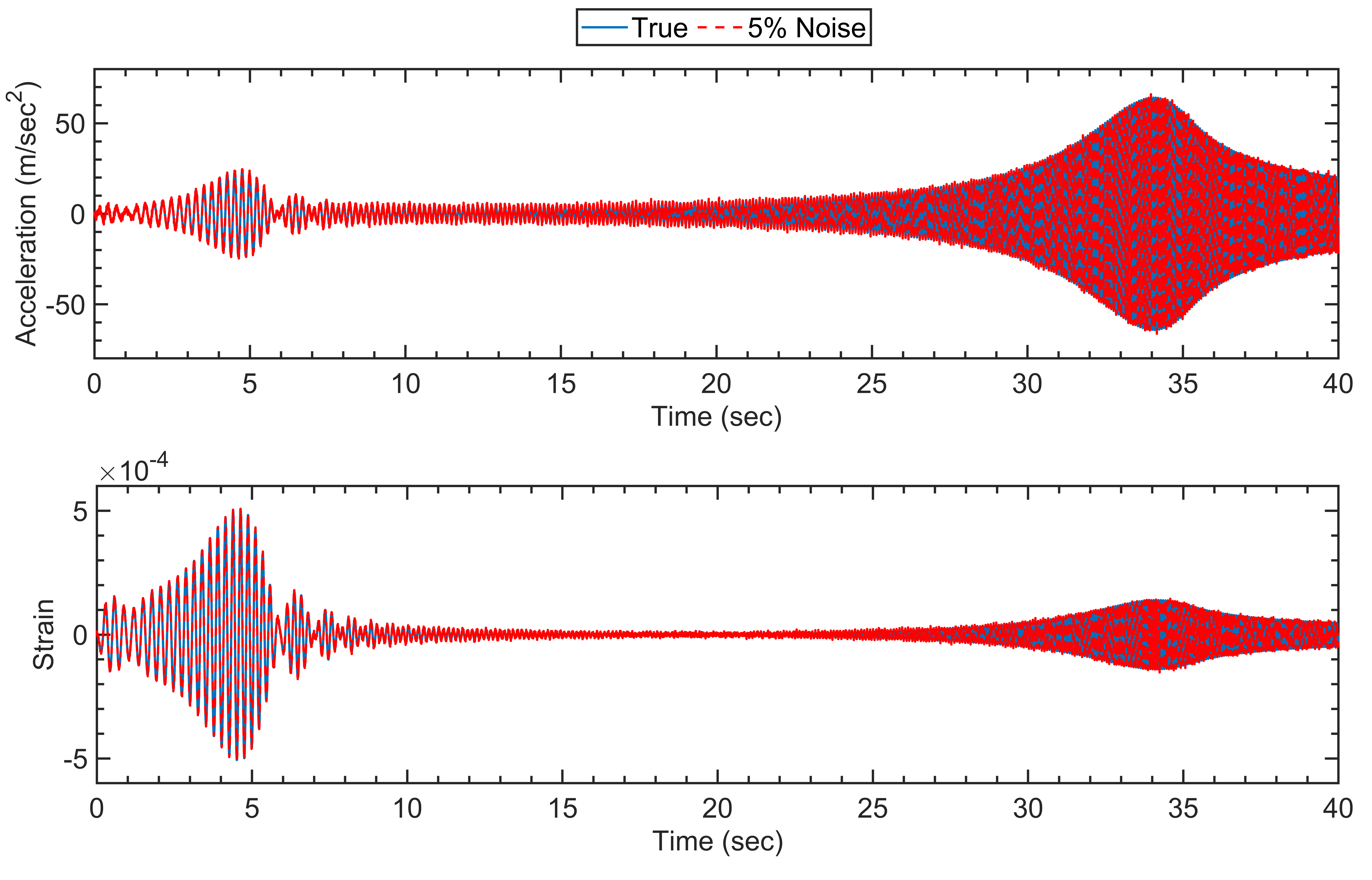}
    \caption{Acceleration and strain data from FE model at x = 0.75 m}
    \label{fig:AccStrainFromFEM}
\end{figure}

The FE model of the tapered cantilever beam is generated in MATLAB by dividing the length $L$ of the beam into $n_{e} (= n-1 = 110)$ equally spaced linear elements where $n (= 111)$ is the number of nodes. The depth of the beam at each node is given by $\begin{Bmatrix} h(1) & h(2) & ... & h(n) \end{Bmatrix}$. The dynamic displacement time history of the beam was recorded at the nodes, along with the acceleration and strain data at the corresponding sensor locations for the application of data fusion. The acceleration and the strain signal at $x=0.75 \text{ m}$ (both noisy and true signals) have been shown in Figure \ref{fig:AccStrainFromFEM}.

\begin{figure}
    \centering
    \includegraphics[width=1.0\linewidth]{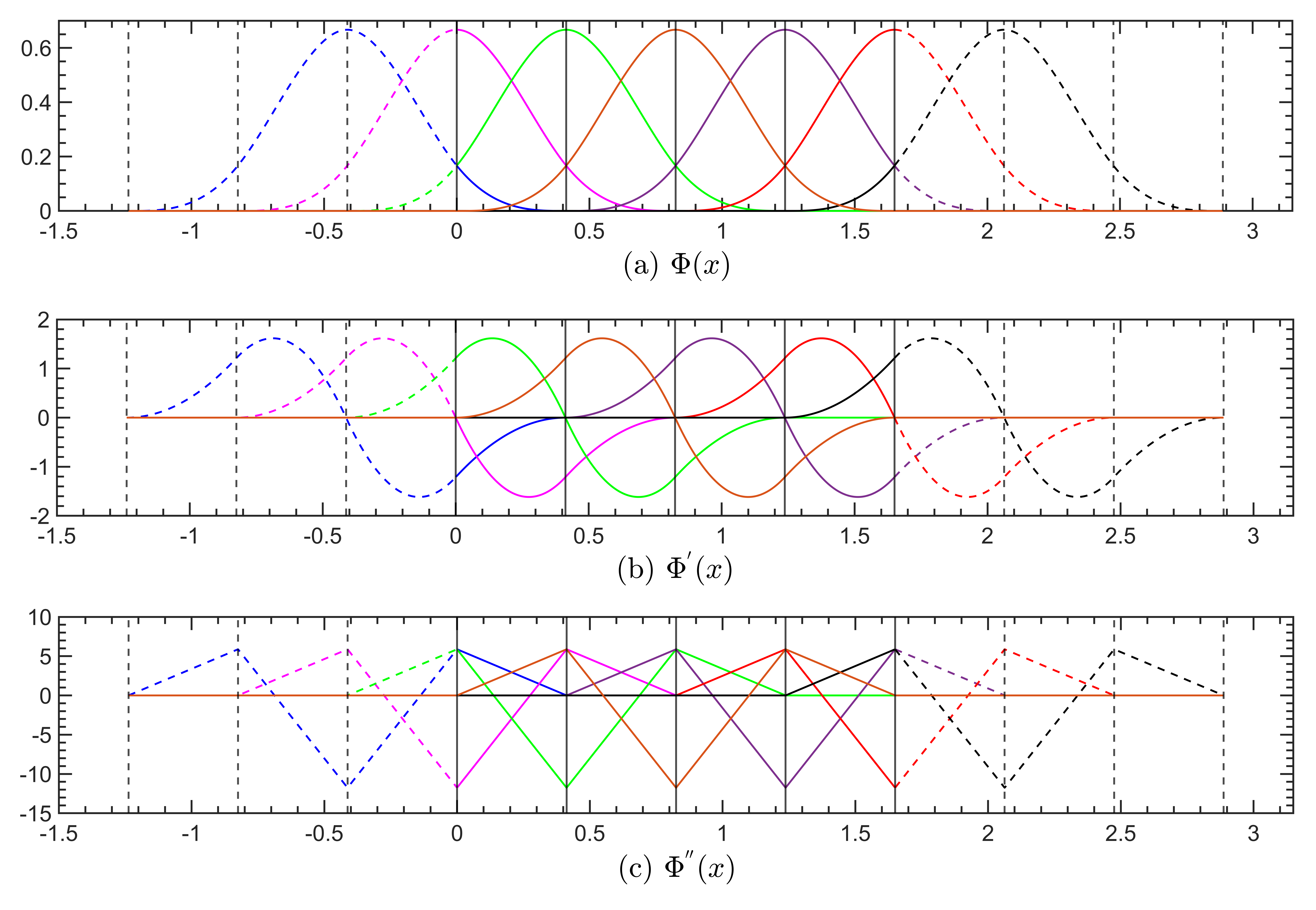}
    \caption{B-spline basis functions and its first and second derivatives with full support at the boundaries}
    \label{fig:NumValSplines}
\end{figure}

B-spline functions of degree $N = 3$ (order $o = 4$) have been considered for applying the data fusion technique in this problem. Since the spatial derivative of degree two of the B-spline functions is involved in the formulation, we need to use closed-ended splines to ensure a smooth second derivative of the basis functions at the boundaries. However, closed-ended B-splines do not have full support at the boundary as each basis function of order $o$ requires $(o+1)$ number of knots to be defined. Hence, the knot vector for this problem was constructed by considering $5$ equally spaced knots inside the domain $[0,1.65]$, including the boundaries, and $N = 3$ external knots on each side of the extended boundaries to get full support. A total $m \; (=k-o-1=7)$ B-spline basis functions are constructed using $k=11$ knots, as shown in Figure \ref{fig:NumValSplines}. Also, we have shown the first and the second derivatives of the spline curves (also used in formulation in Section \ref{sec:Theory}) in Figure \ref{fig:NumValSplines}; here, the internal knots have been marked by solid vertical lines and external knots by dotted vertical lines.

\begin{figure}
    \centering
    \includegraphics[width=1.0\linewidth]{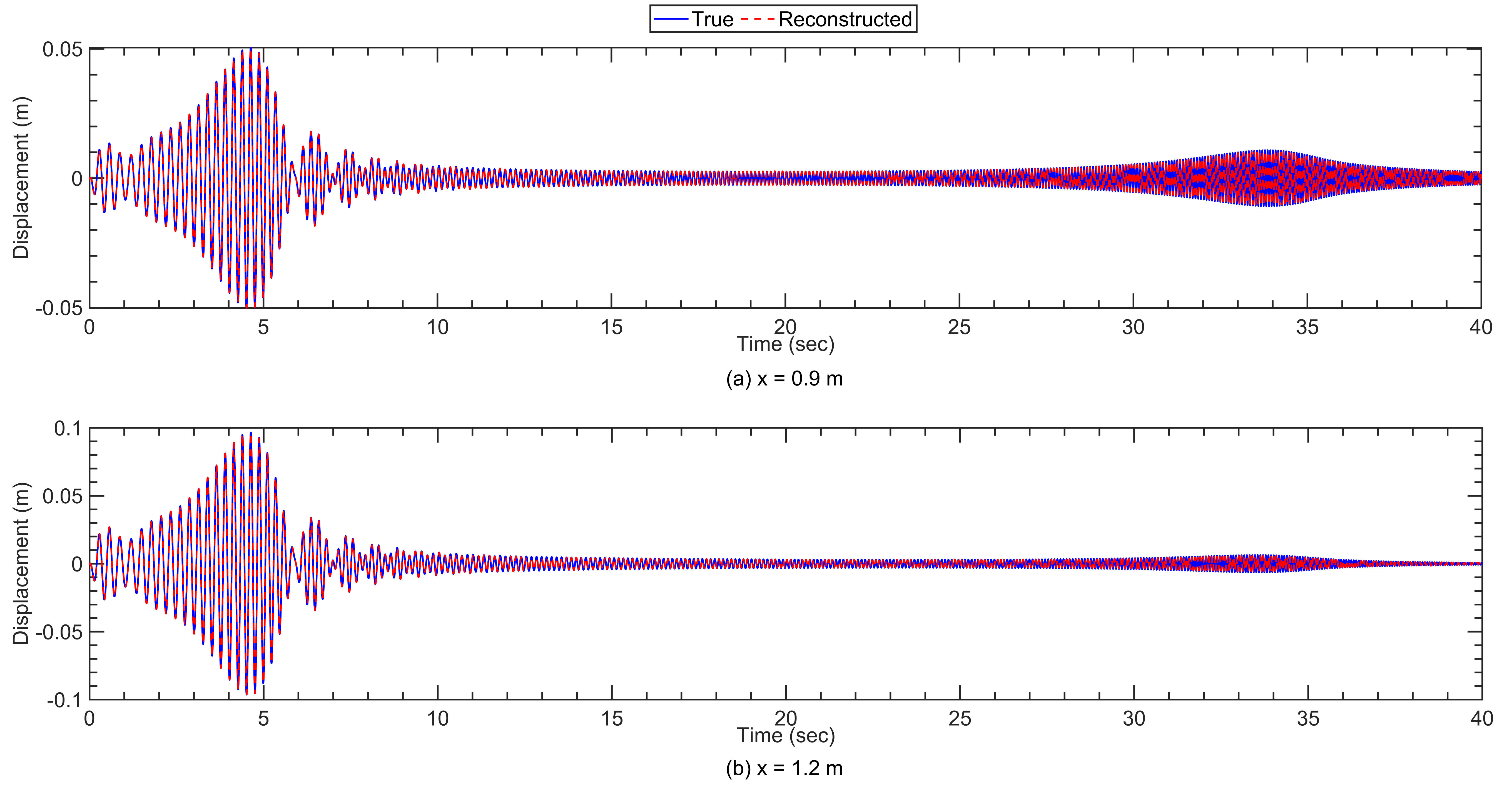}
    \caption{Comparison of reconstructed and true displacements for collocated sensor layout}
    \label{fig:ColloatedDispNumVal}
\end{figure}

\begin{figure}
    \centering
    \includegraphics[width=1.0\linewidth]{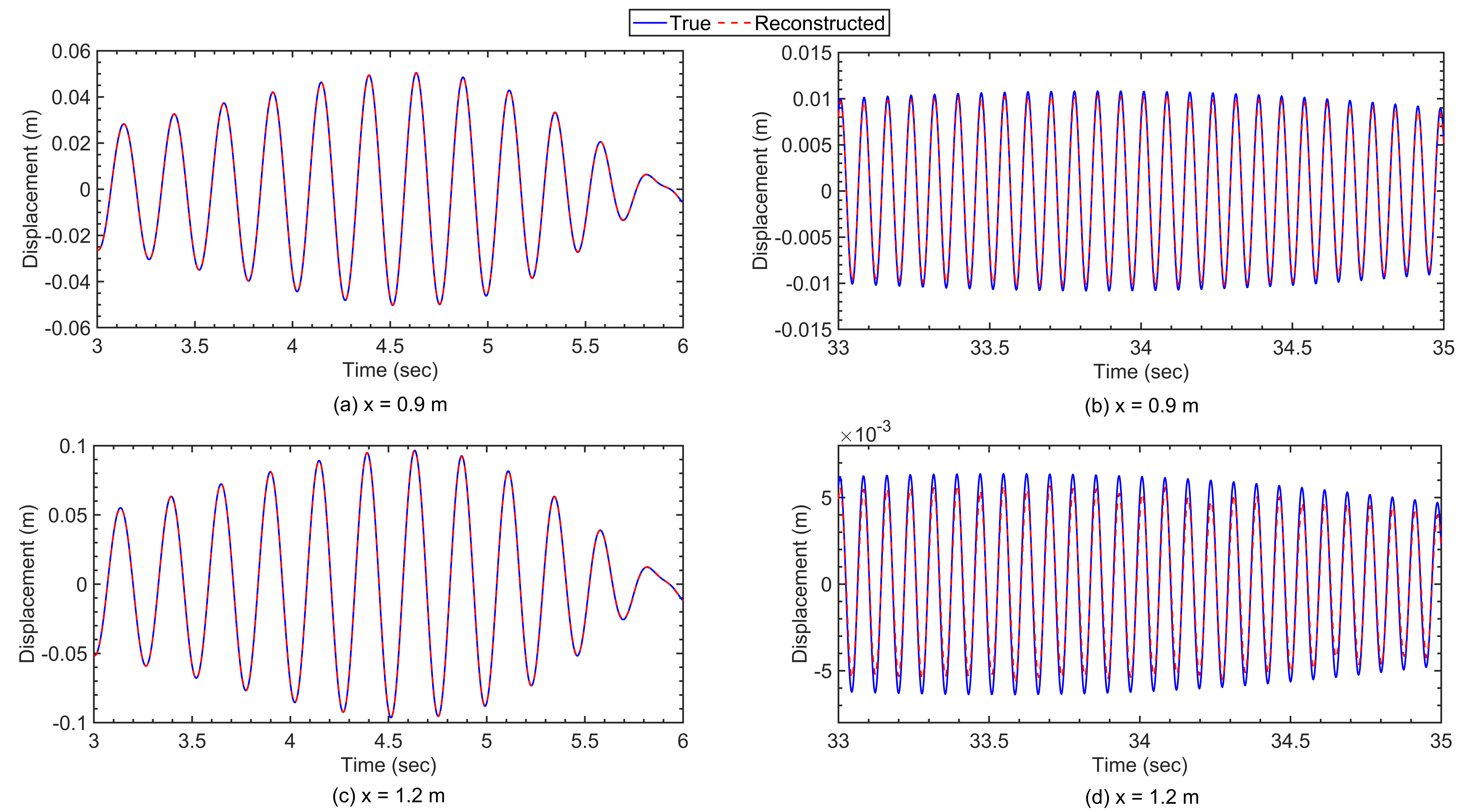}
    \caption{Reconstructed vs true displacements enlarged at the first two modes}
    \label{fig:ColloatedEnlargedDispNumVal}
\end{figure}

From the FE response, a total of $p=8$ acceleration signals and $q=8$ strain signals are recorded, and $5\%$ noise is added to each one of them to mimic experimental conditions. The proposed data fusion technique was applied to the acceleration and strain data to estimate the dynamic displacement. Then, the fused displacement signals were compared with the true displacements from the FE model (as shown in Figure \ref{fig:ColloatedDispNumVal}). The time history of the reconstructed and true displacements have been compared at two locations, i.e., at $x=0.9 \text{ m}$ and $x=1.2 \text{ m}$, along the length of the beam. The two prominent peaks in the displacement time history correspond to the first two flexural modes of the cantilever beam. The reconstructed displacement is observed to match well with the true system response, even in between the two modes where the magnitude of the displacement signal is significantly lower than that near the modes. In Figure \ref{fig:ColloatedEnlargedDispNumVal}, the displacement comparisons have been shown for the two locations more closely at $t=3-6 \text{ sec}$ and $t=33-35 \text{ sec}$, corresponding to the first and second flexure modes of the system. The proposed technique is observed to capture the dynamic displacement well at each of the two modes where the contribution from one of the nodes is more significant than the other.

\subsection{Data-fusion for non-collocated sensor layout}
\label{sec:NumVaildNonCollocatedSensors}
\begin{figure}
    \centering
    \includegraphics[width=1.0\linewidth]{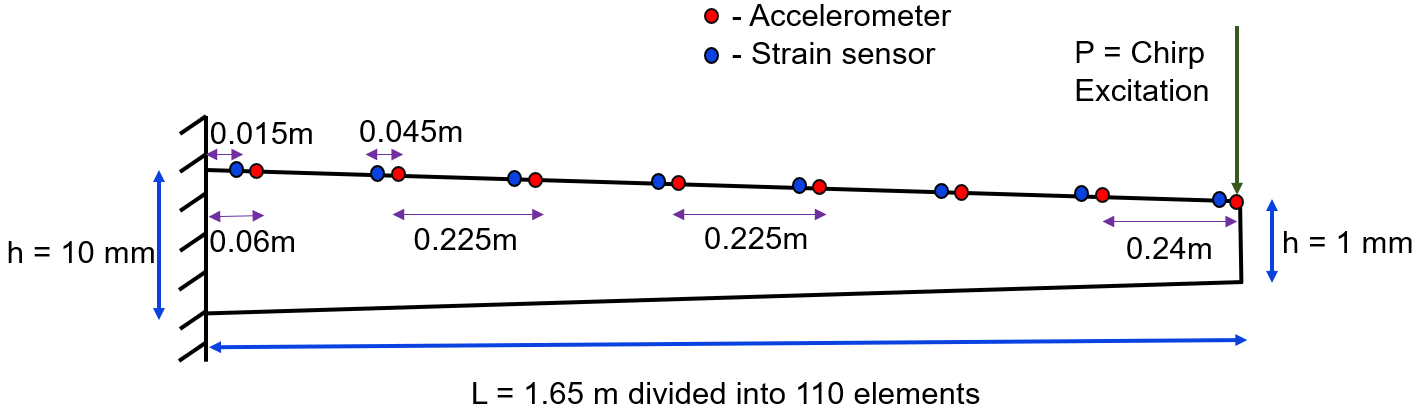}
    \caption{Tapered cantilever beam with non-collocated sensors under chirp excitation}
    \label{fig:NonCollocatedTaperedCantilever}
\end{figure}

In this section, the same tapered cantilever beam fixed at one end has been considered for numerical validation but with non-collocated acceleration and strain sensors. The location of the eight accelerometers and eight strain gauges have been shown in Figure \ref{fig:NonCollocatedTaperedCantilever}. The beam is subjected to chirp excitation at the free end, similar to Section \ref{sec:NumVaildCollocatedSensors}.

\begin{figure}
    \centering
    \includegraphics[width=1.0\linewidth]{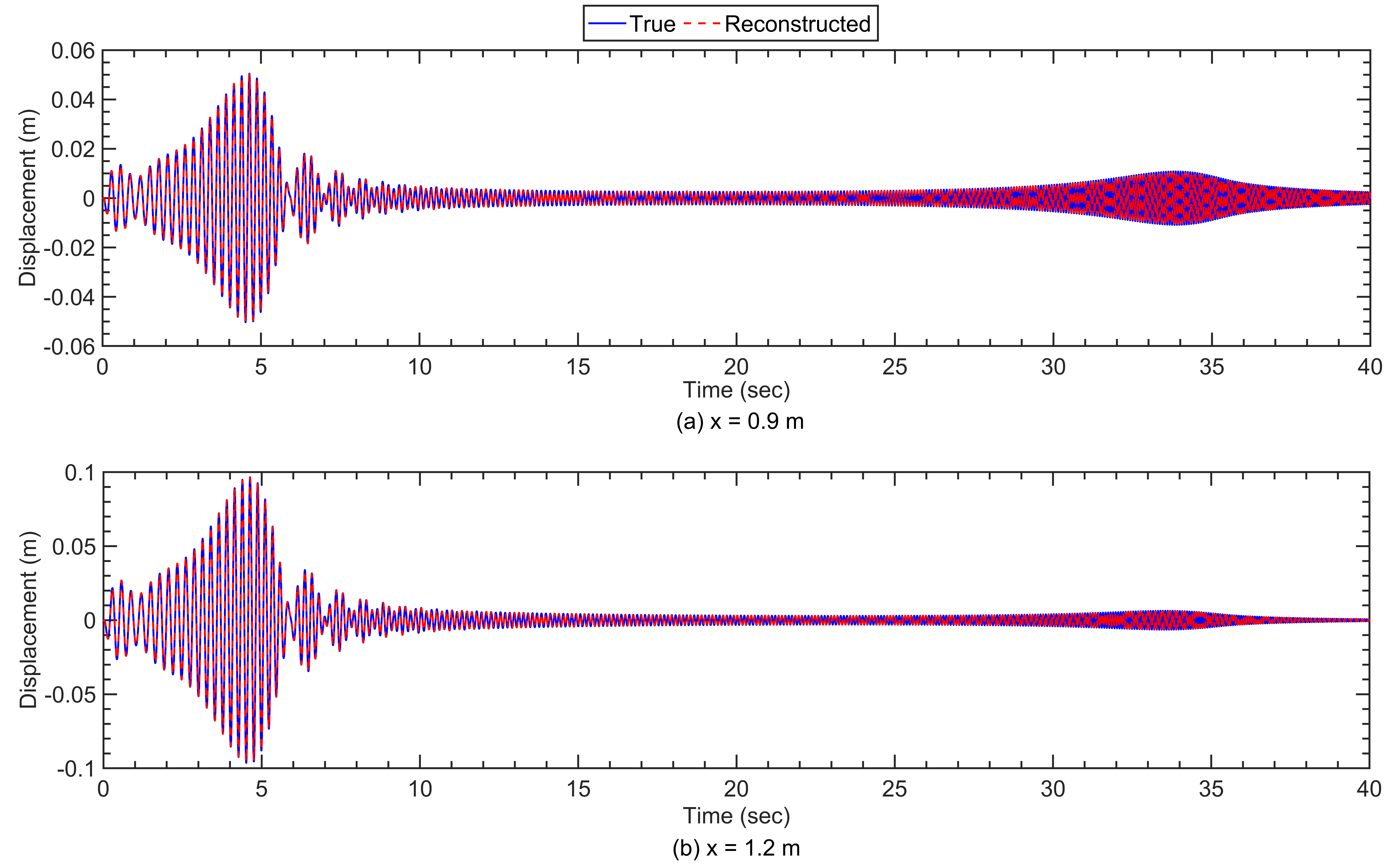}
    \caption{Comparison of reconstructed displacement from data fusion with true displacement}
    \label{fig:NonCollocatedDispNumVal}
\end{figure}

\begin{figure}
    \centering
    \includegraphics[width=1.0\linewidth]{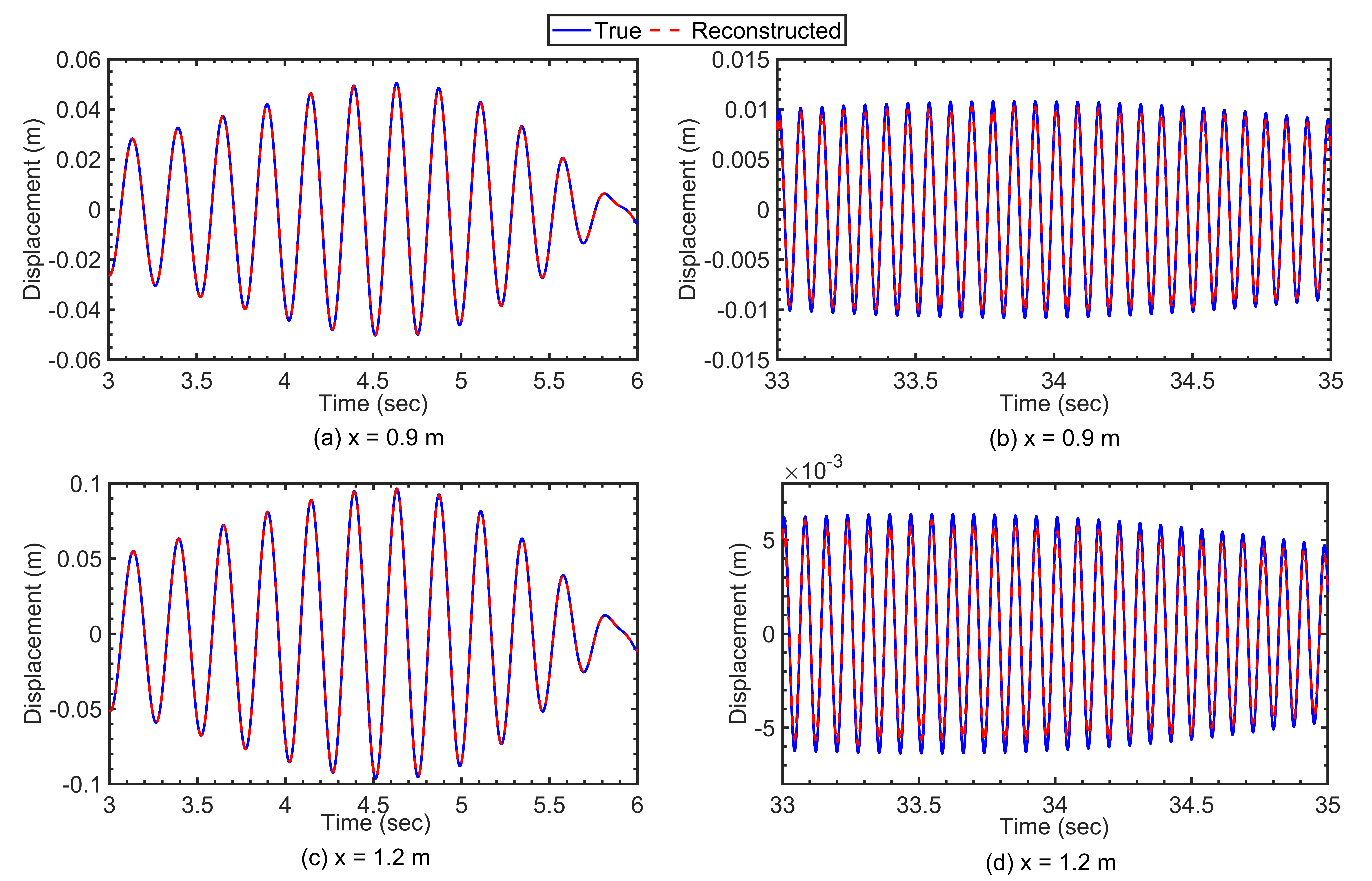}
    \caption{Reconstructed displacement vs true response enlarged near the first two modes}
\label{fig:NonCollocatedEnlargedDispNumVal}
\end{figure}

The acceleration and the strain data are generated at the corresponding sensor locations by simulating the FE model, and $5\%$ noise is added to each of them. The definition for the knot-vector used and the formulation of the B-spline basis functions, in this case, remain the same as in Section \ref{sec:NumVaildCollocatedSensors}. The data fusion technique is applied to the noisy acceleration and strain sensor data, and the dynamic displacement of the beam is estimated. The displacements reconstructed by data fusion at locations $x=0.9 \text{ m}$ and $x=1.2 \text{ m}$ have been compared with the true displacements in Figure \ref{fig:NonCollocatedDispNumVal}. The first two flexure modes are prominent in the displacement time history, and the reconstructed displacement is observed to match the actual response well. Further, in Figure \ref{fig:NonCollocatedEnlargedDispNumVal}, the displacement comparison has been shown more closely for time instants $t=3-6 \text{ sec}$ and $t=33-35 \text{ sec}$ corresponding to the first and second flexure modes respectively. The dynamic displacement of the beam is well-captured by data fusion at the two peaks corresponding to the two flexure modes of the system.

\begin{figure}
    \centering
    \includegraphics[width=1.0\linewidth]{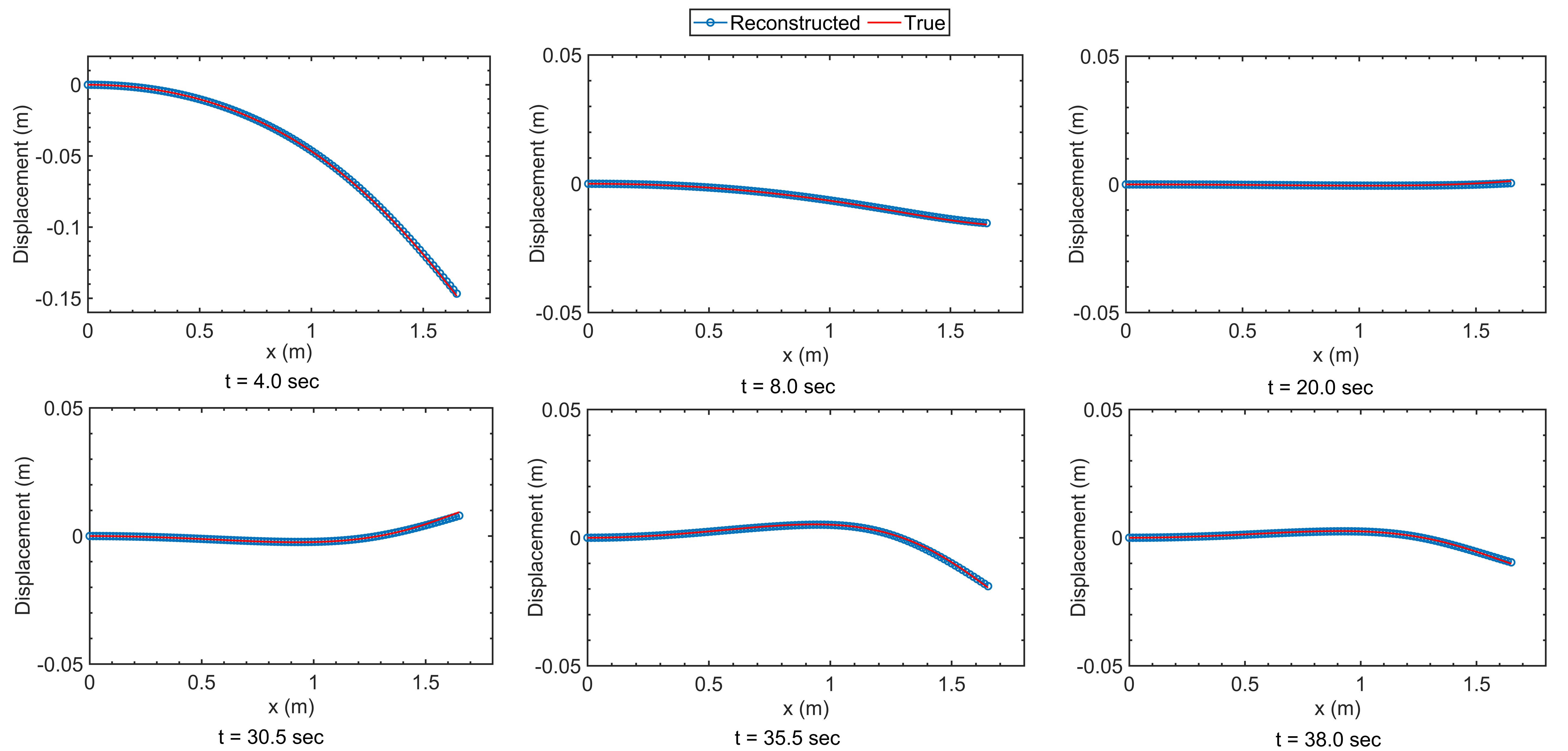}
    \caption{Reconstructed displacement vs true displacement along the length of the beam at different time instants}
    \label{fig:NonCollocatedDispNumValTimeSteps}
\end{figure}

The dynamic displacement profile along the length of the cantilever beam has been plotted at different time instants as shown in Figure \ref{fig:NonCollocatedDispNumValTimeSteps}. It can be observed that the dynamic displacement at all locations along the length of the beam can be tracked simultaneously at each time instant by this data fusion technique. In other words, full-field dynamic displacement estimation of the system can be performed in an online fashion by using the proposed data-fusion technique.

In this section, it has been demonstrated that the proposed data fusion technique for displacement estimation is applicable both for collocated and non-collocated acceleration and strain sensors. Also, the displacement profile of the whole beam can be tracked in an online fashion only with a limited number of sensors.

\subsection{Error Metric}
\label{sec:NumValError}
\begin{figure}
    \centering
    \includegraphics[width=1.0\linewidth]{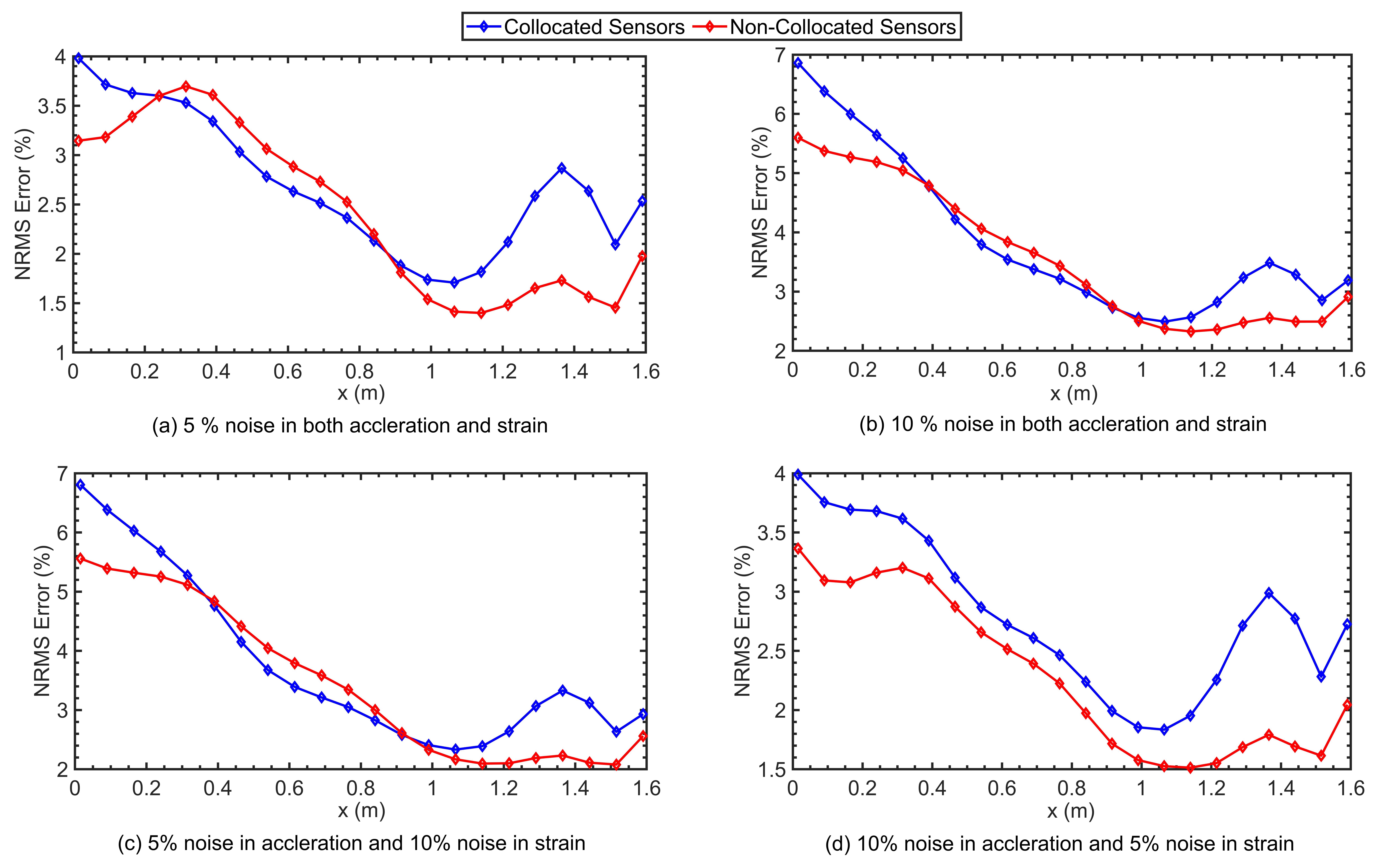}
    \caption{NRMS error in reconstructed displacement along the length of the beam}
    \label{fig:ErrorDisp}
\end{figure}

The error of the displacement estimated by the data fusion technique can be quantified as the normalized root mean square (NRMS) error, as described below:
\begin{equation}
\label{eq:ErrorDef}
    \text{NRMS Error (\%)} = \frac{\text{RMS }{(x_{est}-x_{true})}}{\text{RMS }(x_{true})} \times \text{100},
\end{equation}
where $x_{est}$ is the estimated displacement signal and $x_{true}$ is the true displacement signal at any location along the length of the beam. The numerical validation for the tapered cantilever beam was further carried out with different noise levels in the data. Both collocated and non-collocated sets of sensors were used for noise levels of $5\%$ and $10\%$, and the co-variance matrices $\mathbf{Q_{d}}$ and $\mathbf{R_{d}}$ were tuned accordingly in each case. The NRMS error percentages at different locations along the length of the beam have been plotted in Figure \ref{fig:ErrorDisp}. $5\%$ and $10\%$ noise in both acceleration and strain data have been considered in Figure \ref{fig:ErrorDisp} (a) and (b) respectively; Figure \ref{fig:ErrorDisp} (c) describe the NRMS error for $5\%$ noise in acceleration and $10\%$ noise in strain, and Figure \ref{fig:ErrorDisp} (d) describe the NRMS error for $10\%$ noise in acceleration and $5\%$ noise in strain data.

In both collocated and non-collocated cases, the NRMS error percentage is observed to decrease as the amplitude of the displacement signal increases as we move away from the fixed support. Also, the error increases slightly near the nodal region of the second mode of the cantilever beam, where the displacement is small. Therefore, it can be concluded that the proposed data fusion technique can accurately estimate the full‐field displacement from the noisy acceleration and strain signals.

\subsection{Sensitivity to the number of splines}
\label{sec:SensivityAnalysis}
The selection of the optimum number of splines required for data fusion by the proposed algorithm has been discussed here. The optimum number of basis functions depends on different factors like the dominant modes present in the dynamic response of the system, the number of acceleration and strain sensors, and most importantly, the percentage of noise in the acceleration and strain signals recorded. While defining the spline basis, we need to ensure that we get full support at the domain boundaries and that the number of splines is less or equal to the minimum number of either the acceleration or strain sensors (i.e., $\{(p,q) \geq m\}$ as mentioned in Section \ref{sec:DisFormulation}) to avoid an under-determined system of equations. Increasing the number of splines will help us to capture the contribution of the higher modes of the system to the displacement and thereby increase the accuracy of displacement estimation.

\begin{figure}
    \centering
    \includegraphics[width=1.0\linewidth]{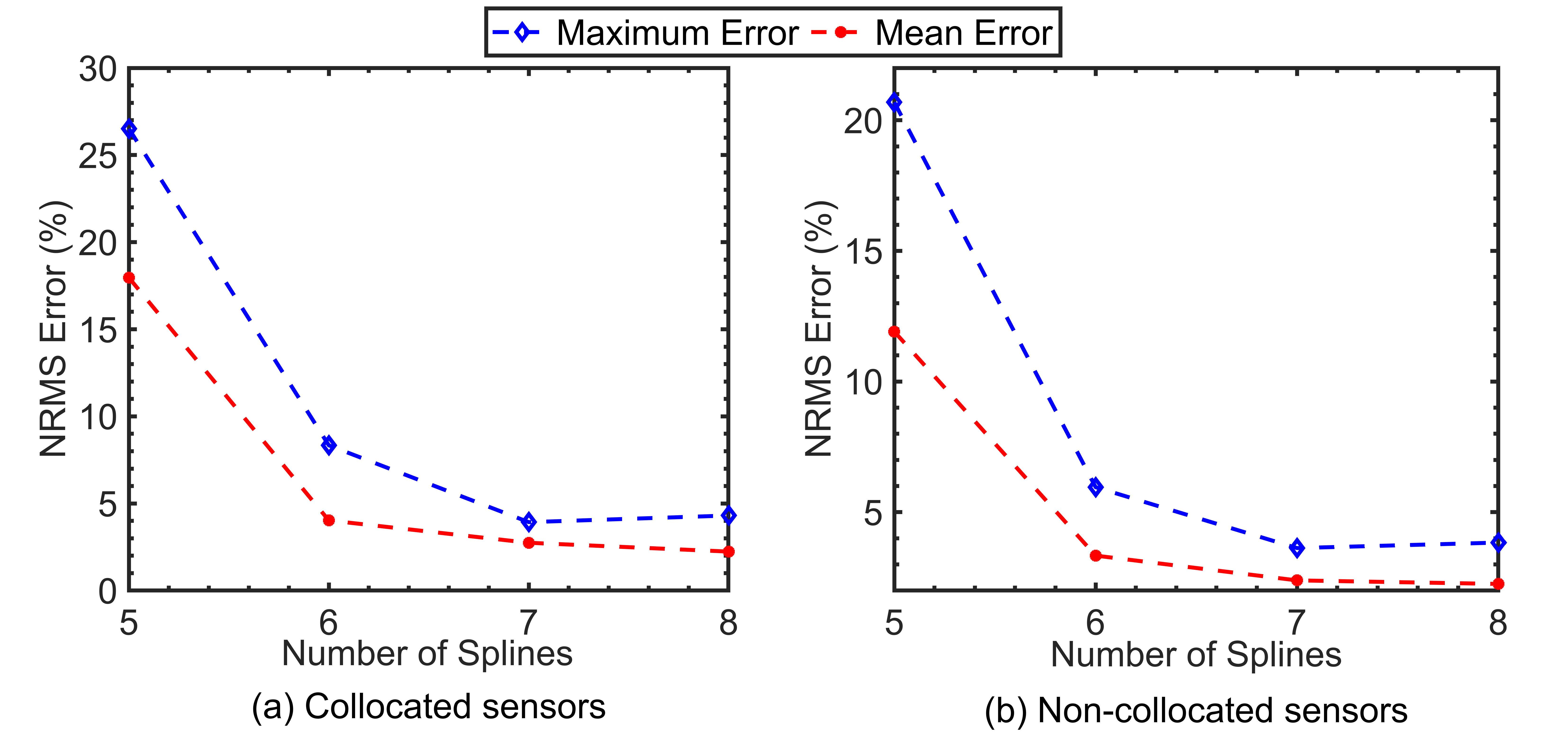}
    \caption{Variation of NRMS Error with number of splines used for data fusion}
    \label{fig:SplineSensitivity}
\end{figure}

In the numerical examples demonstrated in Sections \ref{sec:NumVaildCollocatedSensors} and \ref{sec:NumVaildNonCollocatedSensors}, we need a minimum of six splines (three with non-zero values at each support), while we can select a maximum of eight splines since the number of both acceleration and strain sensors is eight. The variation of the mean and the maximum NRMS Error percentage with the number of basis functions has been plotted in Figure \ref{fig:SplineSensitivity}. The plots show that the optimum number of splines to use in both numerical problems is seven (corresponding to minimum NRMS error).

\section{Experimental Validation}
\label{sec:ExpValidation}
The proposed data fusion technique was experimentally validated using benchmark test data \cite{ou2021vibration} on a small-scale wind turbine (WT) blade subjected to dynamic loading. Experimental benchmark data include acceleration and strain responses measured at various points on a $1.75\text{ m}$ long blade of Sonkyo Energy’s Windspot $3.5\text{ kW}$ WT, tested under dynamic loading with different damage scenarios and environmental conditions in a climatic chamber. The total mass of the blade is $5.0\text{ kg}$, and it is made up of composite materials; more details about the structural properties of the blade are available in \cite{ou2021vibration}.

In analogy to Windspot $3.5$ kW WT, a fix-free set-up was used for the dynamic testing of the turbine blade, where a steel frame was used to fix the blade at one end through four bolts in a temperature and humidity-controlled chamber. The turbine blade was tested both in healthy (case R) and in different damage scenarios (cases A-L) listed in \cite{ou2021vibration}. For each case, the temperature was varied from $-15^{o}\text{C}$ to $+40^{o}\text{C}$, with a step increase of $5^{o}\text{C}$.

\begin{figure}
    \centering
    \includegraphics[width=1.0\linewidth]{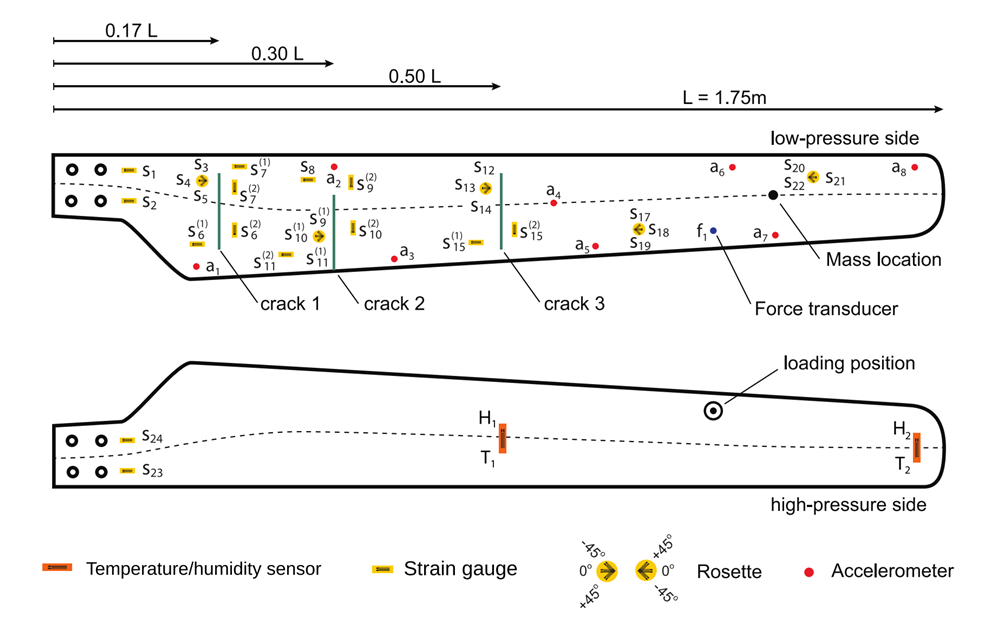}
    \caption{Sensor position layout of wind turbine blade as used in the experiment}
    \label{fig:WTExpSensorLayout}
\end{figure}

In the experimental setup, a combination of eight accelerometers and $18$ strain gauges (consisting of unidirectional strain gauges and rosettes) were installed on the low-pressure side of the blade, along with a force transducer, to track the actual input signal. The experiment was conducted with two different layouts of the strain gauges denoted by superscript $1 \text{ and } 2$ in Figure \ref{fig:WTExpSensorLayout}. For the experimental verification of the proposed data fusion technique, the experiment with the healthy blade condition with sensor layout $1$ (denoted by $\text{Case}\_\text{A}\_(+\text{25})\_1$) under sine-sweep excitation was used. The online data repository was used to obtain the acceleration, strain, and measured input force data corresponding to this specific experiment.

\subsection{Digital twin used for wind turbine simulation}
\label{FEmodelWindTurbine}
The response data of the benchmark experiment \cite{ou2021vibration} was first used to identify the modal parameters of the blade [50]. Then, the distributed structural properties of the blade were obtained by optimizing the composite blade model developed in Numerical Manufacturing And Design (NuMAD), as shown in Figure 15, to minimize errors in modal parameters of the digital twin model compared with identified parameters \cite{SajeerDasEMI2023}. In the experiment, the first $10 \text{ cm}$ of the $1.75 \text{ m}$ length of the wind turbine blade was used to clamp it to the steel support frame through the bolts. Hence, the FE model of the blade was generated by dividing the remaining $1.65 \text{ m}$ length of the blade into $32$ linear elements. The data fusion technique proposed in this article was first applied to the acceleration and strain data from the experiment to get the displacement time history of the blade. Then, it was validated against the analytically obtained displacement response from the FE model.
\begin{figure}
    \centering
    \includegraphics[width=0.8\linewidth]{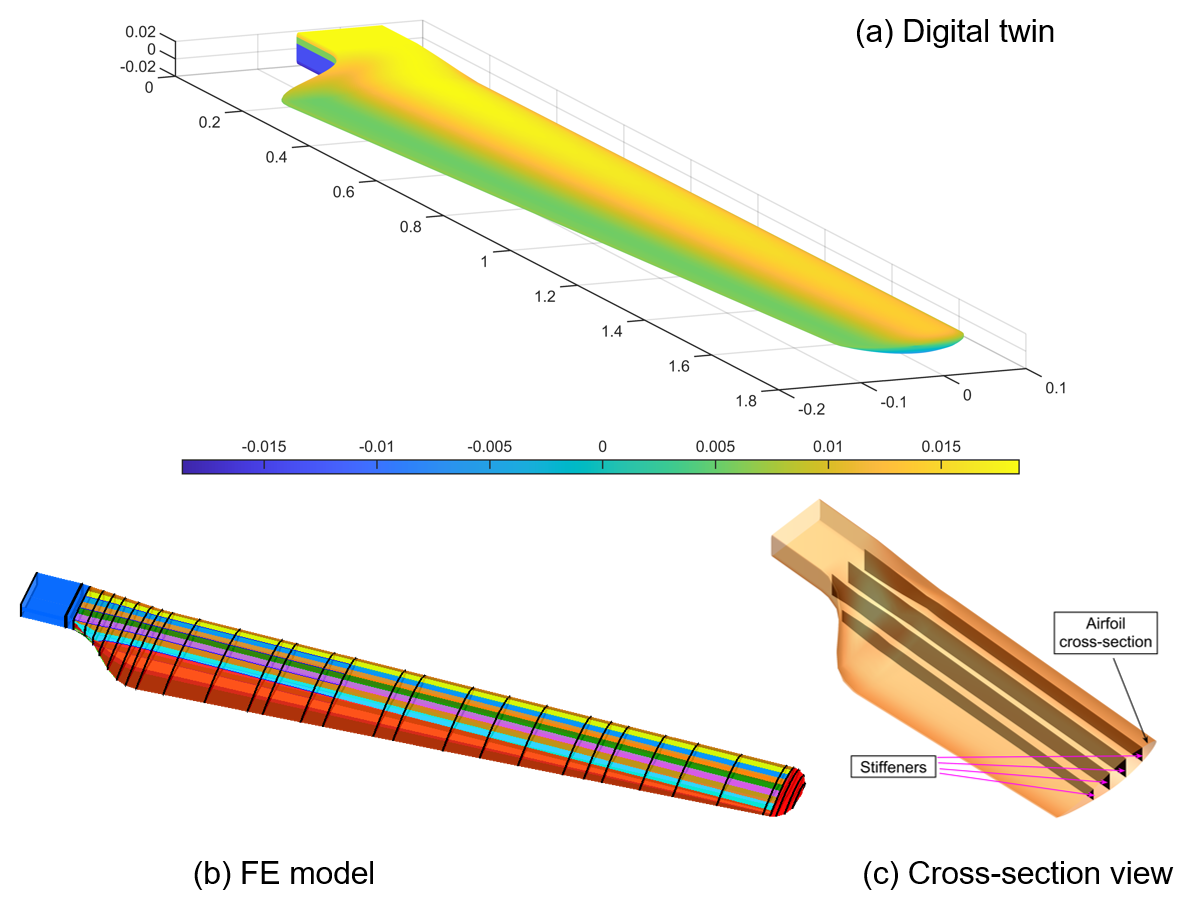}
    \caption{Composite wind turbine blade model developed in NuMAD to obtain distributed structural properties}
    \label{fig:WTFEmodel}
\end{figure}

\subsection{Validation results}
\label{ExpValResults}
The data from all eight accelerometers and the strain gauges oriented along the $z$ axis and associated with flexural strain were selected for the experimental validation. The ten strain gauges selected from the $\text{Case}\_\text{A}\_(+\text{25})\_1$ dataset are $S_{2} \text{, } S_{4}\text{, } S_{6}^{1} \text{, } S_{7}^{1} \text{, } S_{8} \text{, } S_{10}^{1} \text{, } S_{13} \text{,} \\ S_{15}^{1} \text{, } S_{18} \text{, and } S_{21}$, according to the sensor locations and orientation from Figure \ref{fig:WTExpSensorLayout}. The input force data, measured by the transducer, has been provided in the dataset but is highly contaminated by the system response, particularly near the natural frequency modes of the system. Therefore, it could not be used as the true input force for the simulation of the FE model. Instead, a sine-sweep of frequency linearly varying from 1-300 Hz (as was used in the experiment \cite{ou2021vibration}) has been used here to generate the analytical response.

\begin{figure}
    \centering
    \includegraphics[width=1.0\linewidth]{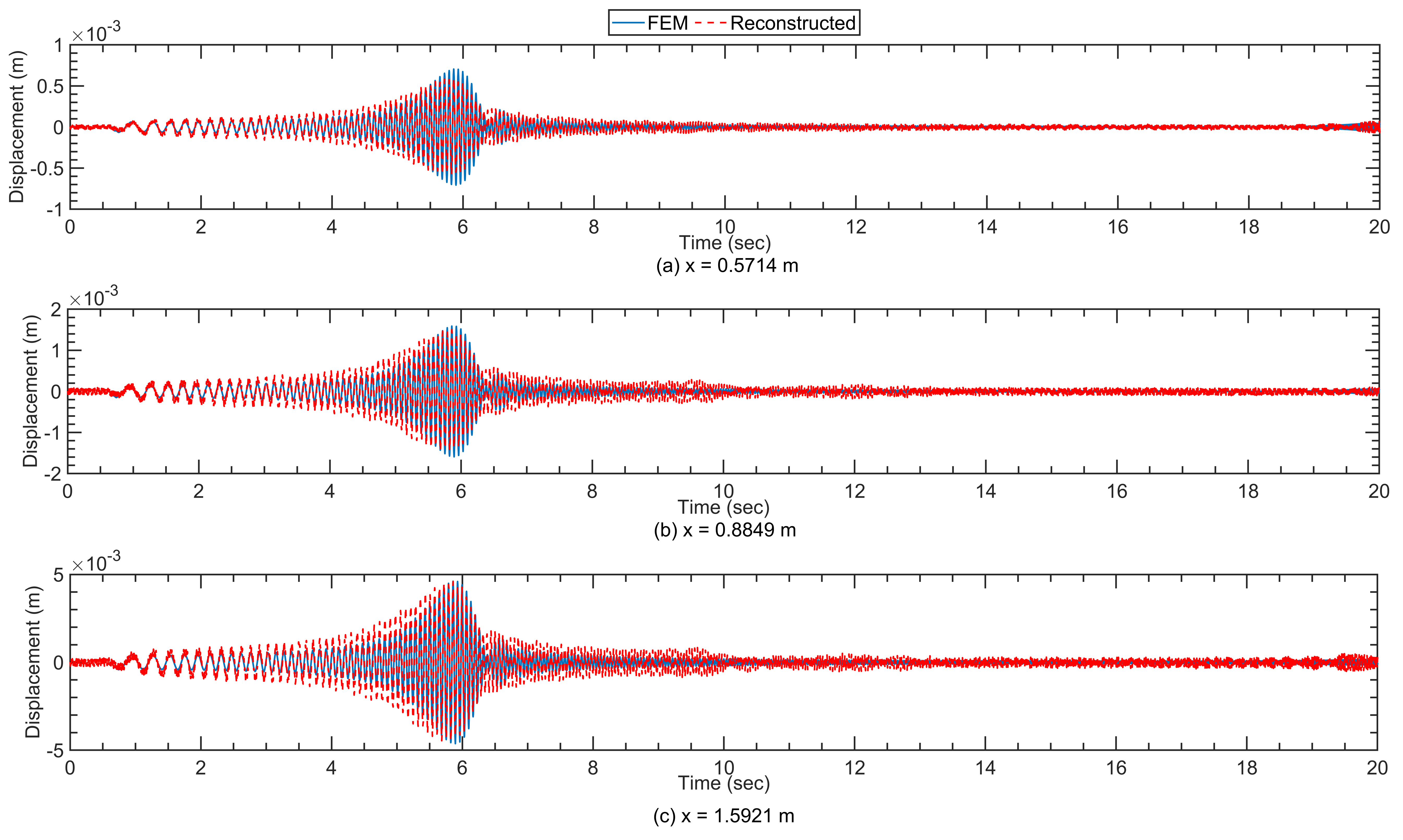}
    \caption{Displacement from data fusion of experimental data vs FE model response}
    \label{fig:WTDispExpVal}
\end{figure}

\begin{figure}
    \centering
    \includegraphics[width=1.0\linewidth]{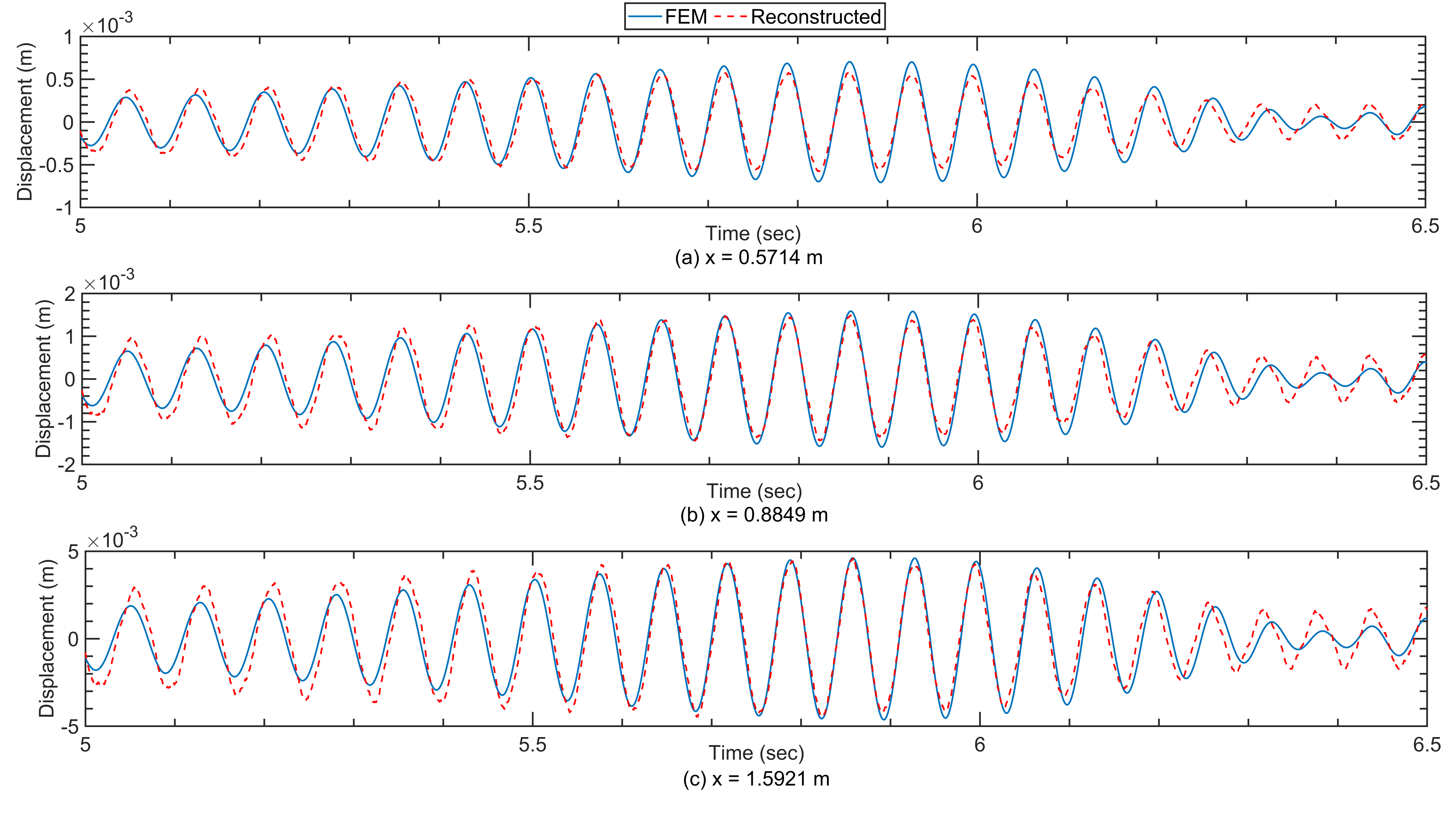}
    \caption{Experimental validation of data fusion enlarged at the first mode}
    \label{fig:WTEnlargedDispExpVal}
\end{figure}

For the application of data fusion for the wind turbine blade, a similar type of B-spline basis formulation was used as in Section \ref{sec:NumVaildCollocatedSensors}. It consists of B-spline functions of degree $3$ (or order $4$), constructed with $11$ equally spaced knots. Then, the proposed data fusion was applied to the wind turbine blade with the experimentally obtained acceleration and strain data. The displacement estimated by data fusion has been compared with the system response generated from the FE model at $x=0.571 \text{ m}$, $x = 0.885 \text{ m}$, and $1.592 \text{ m}$ in Figure \ref{fig:WTDispExpVal}. The reconstructed displacement is observed to match the FE response well of the WT blade. The displacement profiles of the turbine blade have been shown closely for $t = 4.5 - 6.5 \text{ sec}$ in Figure \ref{fig:WTEnlargedDispExpVal} corresponding to the first flap-wise mode.

In the FE model considered here, the blade has been modeled as a cantilever beam under flexure with two degrees of freedom, viz., the vertical displacement in $x$ direction and the rotation about $y$ direction at each node (as shown in Figure \ref{fig:WTFEmodel}). Therefore, the model could capture only the flap-wise modes of the blade, while the experimentally obtained data will have the contributions the flap-wise, edgewise, and torsional modes along with the coupled modes \cite{tatsis2021vibration}. Hence, some differences can be observed in Figure \ref{fig:WTDispExpVal} between the FE response and the displacement reconstructed from the experimental data. However, the reconstructed displacement is observed to match the FE response quite well near the first flap-wise mode, as shown in Figure \ref{fig:WTEnlargedDispExpVal}.

\section{Summary and Conclusion}
\label{sec:Conclusion}
This paper proposes a novel heterogeneous data fusion technique for full-field displacement estimation using only a limited number of sensors. B-spline basis functions have been used to formulate the data fusion algorithm for online displacement estimation in dynamic systems. The proposed Kalman filter-based algorithm fuses experimentally measured acceleration and strain signals to reconstruct dynamic displacement. The use of B-spline basis functions provides an alternative to the use of system information like mode shapes and FE models of the structural system. For this reason, the proposed method is solely based on signal processing and only uses basic problem-specific system information like boundary conditions and the distance of the neutral axis from the location of strain sensors. Therefore, it can be claimed that the proposed method is generalized and is applicable even for very complex structures with little system knowledge.

The numerical validation of the proposed method has been demonstrated for a tapered cantilever beam problem in Section \ref{sec:NumericalValidation}. Here, the application of the data fusion technique has been shown for two different layouts of acceleration and strain sensors. The proposed method successfully eliminates the noise in the acceleration and strain data, and the exact match between the reconstructed and numerically generated displacement can be observed. It can be concluded that the method works well for both collocated and non-collocated sensors, and thereby further increasing the versatility of the technique. Also, full-field displacement estimation along the length of the beam has been demonstrated using only a limited number of sensors.

In Section \ref{sec:ExpValidation}, the experimental validation of the proposed technique has been shown based on an experimental benchmark test of a small-scale wind turbine \cite{ou2021vibration} under dynamic loading conditions. It can be observed that the reconstructed displacement from the experimental acceleration and strain matches the FE-simulated response well, thus proving the practicality of the proposed data fusion algorithm.

\section{Conflict of Interest}
The authors declare no potential conflict of interest.

\section{Funding Information}
The funding for this research from the Rice-IIT-K Collaboration Initiation Grant 2023-24 is gratefully acknowledged.


\bibliographystyle{unsrt}  
\bibliography{DataFusionRefs}






\end{document}